\documentclass{article}
\usepackage{subfig}
\usepackage{amsmath}
\usepackage{physics}
\usepackage{amsfonts}

\usepackage{arxiv}

\usepackage[utf8]{inputenc} 
\usepackage[T1]{fontenc}    
\usepackage{hyperref}       
\usepackage{url}            
\usepackage{booktabs}       
\usepackage{amsfonts}       
\usepackage{nicefrac}       
\usepackage{microtype}      
\usepackage{lipsum}		
\usepackage{graphicx}
\usepackage{natbib}
\usepackage{doi}

\title{On the breathing modes at the interfaces of two medium leading to radiation and guiding of spin waves}

\author{ \href{https://orcid.org/0009-0000-2666-1577}{\includegraphics[scale=0.06]{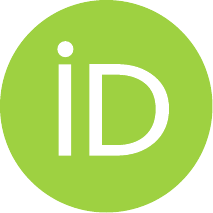}\hspace{1mm}Pushkar Jha}\\
	Department of Electrical and Computer Engineering, Tufts University\\
	Massahusetts, USA  \\
	\texttt{pushkar.jha@tufts.edu} \\
}

\renewcommand{\shorttitle}{\textit{arXiv} Template}

\hypersetup{
pdftitle={A template for the arxiv style},
pdfsubject={q-bio.NC, q-bio.QM},
pdfauthor={David S.~Hippocampus, Elias D.~Striatum},
pdfkeywords={First keyword, Second keyword, More},
}

\begin{document}
\maketitle


\section{Introduction}
Terahertz (THz) radiation, which lies in the frequency gap between infrared and microwaves, typically referred to as frequencies from 100 GHz to 30 THz, has long been studied due to its potential applications in imaging, spectroscopy, biomedical sciences, and integrated circuits \cite{2007NaPho...1...97T, Murphy, Chen_2021, xxxx}. For these applications, THz plasmonic components, e.g., waveguides, based on surface plasmons polaritons have been proposed due to the sub-wavelength confinement for miniaturized devices \cite{article, Maier, Ozb, Schuller2010PlasmonicsFE}. However, these plasmonic waveguides are two-way waveguides; i.e., light waves propagate in both the forward and the backward directions. One-way-propagating waveguides are of particular importance for functional devices such as isolators, switches and splitters\cite{article5}.

Throughout the years, ferrites have evolved as the irreplaceable materials in microwave engineering, mainly due to the off diagonal terms in their permeability tensor and large value of saturation magnetization\cite{book, Sugimoto, article4}. The off diagonal terms leads to the non-reciprocity in phase of the propagating waves in such media. Ferrites non-reciprocal use provides high isolation, good power handling and wide bandwidth for normal use. However, for higher frequencies in the THz range, ferrites start to suffer from drawbacks such as high loss and limited saturation magnetisation \cite{6891475}. Hence, there have been many attempts to utilise other materials that exhibit similar gyrotropic behaviour at higher frequencies to design nonreciprocal components.

Magnetically biased semiconductors such as InSb exhibit significant nonreciprocal gyroelectric (which are the electromagnetic dual of ferrite’s gyromagnetic
properties) behavior at THz frequency. The first semiconductor junction circulator based on InSb was proposed by Davis and Sloan \cite{276774}. Since then, many researchers have investigated the potentials of magnetically biased InSb to design isolators and circulators \cite{925497, 1266872, Jawad2015MillimetreWS, 7836295}. Reported outcomes show promising behaviour despite some challenges like high loss at room temperature and high magnetic bias requirements.
    
These modes can also lead to the radiation of electromagnetic waves \cite{1137884}. Let us consider the case of a dipole antenna placed in vacuum. The near field region of the dipole antenna which is a vacuum lies at the interface of this dipole. As soon we excite the dipole with a electrical current, it starts giving this energy to the medium surrounding it in the form of magnetic energy. This energy keeps on oscillating between this coupled mode form of a magnon and a photon. The photonic spin of the magnons is perfectly locked with that of a photon. This region exists at the immediate vicinity of the antenna and is the actual source of radiation. There have been reports of the presence of causal surfaces in this region which results due the partial standing wave nature of this mode. The magnon-photon system oscillates between two fixed points. The two interfaces formed have capacitive and inductive energy in the coupled state. This region is also called reactive because of the nature of this mode of propagation where energy is purely imaginary. The understanding of this underlying physics will be also indispensable in the development of future radiating structures.

In this paper, we have analyzed the photonic spin of the propagating wave in the gyroelectric waveguide . We have shown how these waves breaks spin-momentum locking and hence are unidirectional. We have also shown the role of gyrotropy in the emergence of breathing and hyperbolic modes and the cut-off modes in such medium. The study of the near field region of the hertzian dipole and a loop antenna reveals the presence of causal surfaces in the near field region of the antenna. Here, we have used Method of Moments to calculate the standing points near the antenna. We have also calculated impedance in the near field region which suggests its imaginary nature in the near field.

\section{Overview of Spin waves}
The theory of wave propagation in a gyrotropic media has been studied extensively in literature. The gyroelectric material can be modelled as a electron plasma, a quasi-neutral gas of charged particles containing electrons and positive ions showing collective behavior. Here, quasi neutrality means that the overall charge densities of the charged particles cancel each other in equilibrium. Collective behavior means that the local disturbances in equilibrium can have strong influence in the remote regions of the plasma but this effect is cancelled out as the size of the plasma becomes greater than the parameter defined for plasmas, called the debye length.

Lets suppose that a positively charged particle +Q is introduced in a plasma. This paricle will attract the free electrons around it and repel the ions. This will introduce a electric field in the region, which can be written as the gradient of scalar potential as,

\begin{equation}
    E = -\nabla\phi
\end{equation} 

This implies that for any electron or ion around the charged particle, we can write their potential energies as $-e\phi$ and $+e\phi$ respectively. Now, according to Boltzmann's law, electron will have tendency to go to regions of low potential energies. This implies that the electron density in the surrounding region can be wriiten as:

\begin{equation}
   n_e = n_oe^{\frac{+e\phi}{k_bT}}
\end{equation}

For $\frac{e\phi}{k_bT} << 1$, this can be written as 

\begin{equation}
   n_e = n_o(1+\frac{e\phi}{k_bT})
\end{equation}

Similarly, the ion density can be written as:

\begin{equation}
   n_i = n_o(1-\frac{e\phi}{k_bT})
\end{equation} 

where, $n_o$ is the density of electron far away from such region and T is the temperature. Now, we can write the poisson's equation in the region around the charge +Q as:

\begin{equation}
   \epsilon\nabla\cdot E = e(n_i - n_e)
\end{equation}

Using (1), this can be written as:

\begin{equation}
   \nabla^2\phi = \frac{e}{\epsilon}(n_e - n_i)
\end{equation} 

Using (3) and (4), we get:

\begin{equation}
   \nabla^2\phi = \frac{2n_oe^2}{\epsilon k_bT}\phi
\end{equation} 

By solving this we get the potential of charge +Q as:

\begin{equation}
   \phi = \frac{Q}{4\pi\epsilon r}e^{\frac{-r}{\lambda_d}}
\end{equation}

here, $\lambda_d$ is called the debye length and can be written as :

\begin{equation}
   \lambda_d = \sqrt{(\frac{\epsilon k_bT}{2n_oe^2})}
\end{equation} 

We can see from the above expression that the electrostatic potential introduced by a charge +Q falls very rapidly as the distance from the charge increases beyond the debye length. This implies that the mutual interaction between charges can be ignored if the size of the plasma is greater than the debye length and hence the collective behavior becomes important. Now, we can write the maxwell's equations in a magnetised plasma as:

\begin{equation}
   \nabla\cdot (\epsilon_o \hat{\epsilon}\cdot E) = 0
\end{equation} 

\begin{equation}
   \nabla\cdot B = 0
\end{equation} 

\begin{equation}
  \nabla \times E = i\omega \mu_o  H
\end{equation} 

\begin{equation}
   \nabla \times H = -i\omega \epsilon_o \hat{\epsilon} E
\end{equation} 

Now, lets substitute $\nabla = ik$ in equations (12) and (13), where k is the propagation vector. This gives:

\begin{equation}
   k \times E = \omega \mu_o  H
\end{equation} 

\begin{equation}
  k \times H = -\omega \epsilon_o \hat{\epsilon} E
\end{equation} 

Combining (14) and (15), we can arrive at the equation:

\begin{equation}
  \hat{D}\cdot E = 0
\end{equation} 

where, $|D| = |k^2\hat{I} - k k - \frac{\omega^2 \hat{\epsilon}}{c^2}| = 0$, is the dispersion relation in the medium. Lets re-write (16) after substituting D:

\begin{equation}
 k^2 E - k k\cdot E = \frac{\omega^2 \hat{\epsilon}}{c^2}\cdot E
\end{equation}

Now, lets assume that the medium is biased along the z direction with homogeneous magnetic field and the k-vector is also along the same direction. The electric field vector will oscillate in the x-y plane. This will make the second term in (17) zero and it will reduce to:

\begin{equation}
 k^2 E = \frac{\omega^2 \hat{\epsilon}}{c^2}\cdot E
\end{equation} 

If we write the x-component of (18), it will give:

\begin{equation}
 (k^2 - \frac{\omega^2 \epsilon_1}{c^2})E_x = (\frac{\omega^2 \kappa}{c^2})E_y
\end{equation} 

Similarly, for y-component, we can write (18) as:

\begin{equation}
(k^2 - \frac{\omega^2 \epsilon_1}{c^2})E_y = (\frac{\omega^2 \kappa}{c^2})E_x
\end{equation}

Using (19) and (20), we can write the dispersion relation in the medium as:

\begin{equation}
(k^2 - \frac{\omega^2 \epsilon_1}{c^2}) = +(\frac{\omega^2 \kappa}{c^2})
\end{equation}

and

\begin{equation}
(k^2 - \frac{\omega^2 \epsilon_1}{c^2}) = -(\frac{\omega^2 \kappa}{c^2})
\end{equation}

This implies that there are two solutions of the propagation constant which is because of the induced gyrotropy due to the magnetic field. This results in finite term in the RHS of (21) and (22). Substituting the value of k that we get from (21) and (22) in (19) and (20), we get two relation between x and y component of the electric field as:

\begin{equation}
E_y = - i E_x
\end{equation} 

and

\begin{equation}
E_y = i E_x
\end{equation} 

This implies that the x-y components of the electric field are 90 degrees out of phase. There are two such waves, one where the y component is leading and hence it is a right circularly polarized light with photonic spin +1. The y component in the other wave is lagging the x-component and hence it is a left circularly polarized light with photonic spin -1.

\subsection{Phononic modes}

The drude model can be applied to get the frequency dependent permittivity tensor in gyroelectric medium \cite{7087406}. The presence of magnetic field leads to non-reciprocal topological isofrequency surfaces such as the ellipsoid and two fold Type I and one fold Type II hyperboloids \cite{Shekhar:19} \cite{Pendharker:18}. The non-reciprocity is due to the existence of the off-diagonal terms in the permittivity tensor in such a media which arises due to the inherent polarization of the electric fields in such a medium which is fixed by the magnetic bias. This has been also referred as the photonic spin of the waves in the literature \cite{Sen_2022}. These propagating waves have their spins locked to the gyrotropic axis of the medium.

It has been observed that different forms of isofrequency surfaces expresses the kind of anisotropy, namely the relations between components of permittivity and/or permeability tensors characterizing the medium. For example, in the isotropic medium these surfaces appear as the closed forms of sphere whereas in the anisotropic medium, they can take both open and closed forms. In the open form, they take the shape of a hyperboloid. In the anisotropic medium, when all the principal value of the permittivity tensor is positive, it takes the closed form while if one or two of them in negative, it takes open form of Type I or Type 2 hyperboloid shown in the Figure \ref{Quadrics} below.

\begin{figure}
\begin{center}
    \includegraphics[scale=.8]{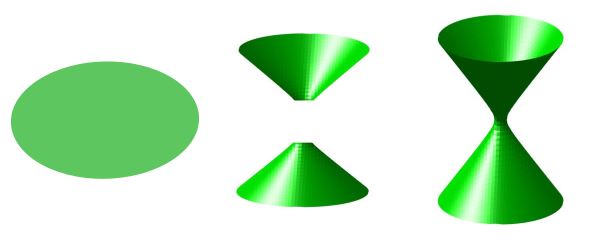}
  \caption{Three typical topologies (ellipsoid, two fold Type I and one fold Type II hyperboloids) of isofrequency surfaces inherent in anisotropic media.}
  \label{Quadrics}
\end{center}
  \label{fig:IsofrequencyContoursRef}
\end{figure}

Despite the fact that hyperbolic topology can be found in natural anisotropic media, there is a considerable interest in creating artificial structures (metamaterials) possessing desired functionality \cite{articlex}. These structures are created by combining together the metallic and dielectric layers into a unified structure. Since, hyperbolicity requires plasma behavior in a certain direction of wave vector space and insulating behavior in the others leads to an option instead of metals along with dielectrics using some combination of semiconductors and magnetic materials (e.g., ferrites) as building blocks of the hyperbolic metamaterials. This text\cite{article1} has demonstrated that depending on the ratio between filling factors of magnetic and semiconductor subsystems within a superlattice and the direction of an external static magnetic field with respect to the structure periodicity, a complex of isofrequency surfaces in the form of a ellipsoid and bi-hyperboloid can be obtained. The bi-hyperboloid isofrequency surfaces has been reported first time in the literature and which is unachievable in ordinary anisotropic medium.

In the absence of gyrotropy, the upper operating frequency boundary for plasmonic system is limited by the plasma frequency of the material that is typically located in the ultraviolet spectrum for noble metals such as gold and silver, but can be located in the far infrared frequencies for highly doped semiconductors such as InSb. The plasma frequency can be also tuned for such materials through doping and/or changing the temperature or the external magnetic fields. 

The presence of propagating waves well below the plasma frequency in the presence of gyrotropy has also been studied in the past. The gyrotropy can lead to the existence of hyperbolic isofrequecy surfaces that cannot be found in absence of gyrotropy. It has been found that these gyrotropy induced waves are fundamentally different from the waves that exists without gyrotropy and breaks the universal spin-momentum locking in real space. The presence of a particular isofrequency surface can be directly attributed to relative magnitude of the diagonal and off-diagonal terms in the permittivity tensor. The increase in gyrotropy can also lead to the emergence of isofrequency surfaces that wouldn't have existed earlier or it can suppress the surface that would have existed. This phenomenon can lead to topologically protected non-reciprocal plasmonic systems at THz frequencies. 

The presence of gyrotropy leads to a frequency bandwidth between the cyclotron frequency (which comes into picture because of the gyrotropy) and the cut-off frequency for the RHCP wave that exists beyond this frequency. This region is called the forbidden region and no propagating electromagnetic waves exists in this region. The width and the location of the region can be tuned by varying the bias magnetic field. This enables the possibility for realizing filters at THz frequencies. There is also a heating phenomena that arises in this region and is of considerable interest. The launching of electromagnetic waves near the cyclotron frequency in the material leads to the heating of the material at certain positions which can be controlled. This happens because the frequency of the waves becomes equal to the cyclotron frequency locally after penetrating some depth inside the material which leads to the deposition of all the electromagnetic energy at such positions. This region supports evanescent waves which gets excited due to this deposition of energy. This phenomenon can also be exploited to realize innovative photonic systems.

The high magnetic bias and very low temperature requirements for InSb are two of the major challenges that needs to be met in order to experimentally realize the designed photonic systems. For THz applications, the minimum dimension for the plasmonic components has to be sufficiently more than few micrometer to satisfy the diffraction limit. This limits the integration of optical integrated circuits with the electronic circuits having dimensions in the nanometer scale. It is clear that InSb cannot be utilized as the gyroelectric medium for dimensions less than a um. We can use metamaterials as a dopant to realize photonic materials. This will enable us to engineer the dielectric tensor of the medium which depends on frequency but these materials itself has to be several wavelengths long, because the typical period is on the order of half of a wavelength\cite{JoannopoulosJohnsonWinnMeade+2011}.

Plasmonics may offer a solution to this size-compatibility issue since it has both the features of photonics and miniturization of electronics by confining the light to very small dimensions which could be in nanometer scale. The sub-wavelength localizations for the propagating plasma oscillations has been achieved using nano-wires but the losses associated with heating within the metallic boundaries limits the maximum propagation length of light within these structures. Thus there is a basic trade-off in all plasmon waveguide geometries between mode size and propagation loss. One can have a low propagation loss at the expense of a large mode size, or a high propagation loss with highly confined light. Compact plasmon waveguides generally suffer from high loss, and chip-scale integration presents a challenge, as does efficient coupling off-chip \cite{Barnes2003SurfacePS}. A hybrid approach, where both plasmonic and dielectric waveguides are used, has been suggested as a solution to this trade-off \cite{Hochberg:04}.

\begin{figure}
    \centering
    \includegraphics[scale=0.5]{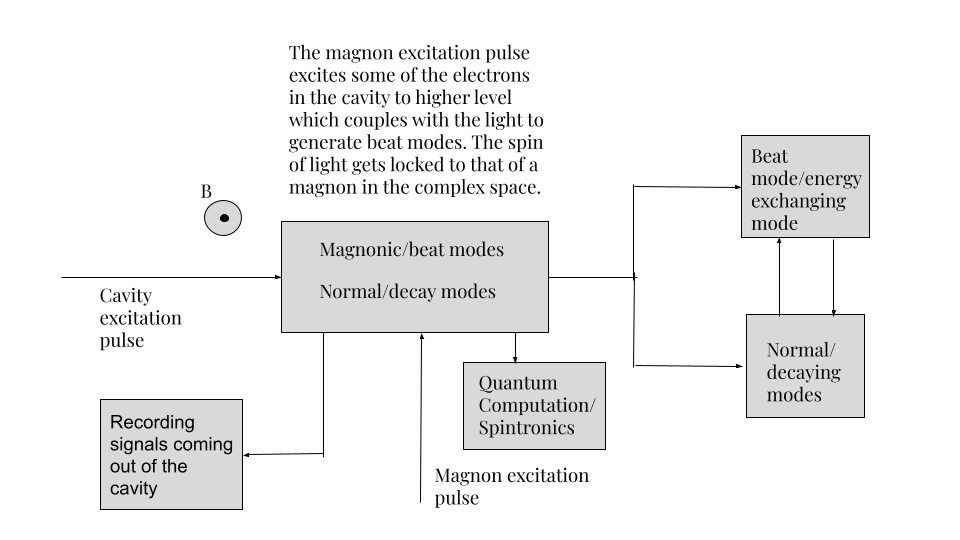}
    \label{Gyrotropic Waveguide}
    \caption{Cavity Dynamics}
\end{figure}

\begin{figure}
    \centering
    \includegraphics[scale = 0.7]{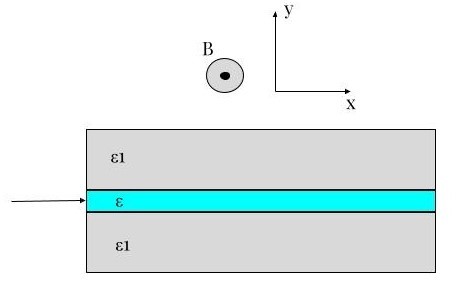}
    \caption{Gyroelectric Waveguide}
    \label{Gyrotropic Waveguidet}
\end{figure}

The surface plasmon polaritons found at the interfaces between different medium can lead to confinement as well as guiding of spin waves that propagates in such structures. It has been reported that such waves results due to the coupling of light with the quanta of vibrational energy in the medium lattice which are called phonons. These modes are oscillatory in nature and breaks time reversal symmetry and spin momentum locking in real space which can lead to one way propagation of large amount of energy at THz frequencies \cite{articlehh}. It has been studied that plasmonics can be the route to merge electronics to photonics at nanoscale dimensions. The introduction of gyrotropy can lead to the cut-off or emergence of such modes in a adjustale frequency window which all can be the building blocks of the future photonic computers \cite{Ghosh2022} \cite{nature}.

In past, many one way propagating plasmonic devices have been proposed based on interference between differently polarised SPPs \cite{2010NatPh...6..126W}. Another approach is to use the nonreciprocal effect of SPPs under an external magnetic field (MF) (also called surface magneto plasmons (SMPs))\cite{PhysRevLett.28.1455}\cite{PhysRevLett.100.023902}. However this device \cite{PhysRevLett.100.023902} is only applicable in the visible frequencies, and the required magnetic bias is too strong to be realized. An alternate approach was proposed where metal-dielectric-semiconductor structure was used (as shown in \cite{Hu:12}\cite{ 6316071}\cite{7836295}) to realize one-way propagation characteristics with magnetic bias of around 1T. If we bias such a structure in the direction perpendicular to the propagation direction, we get different dispersion curves for the forward and backward propagating modes which leads to the non-reciprocal behavior. We also get a frequency band where only one of the modes propagates which can be utilized to realize one-way propagation. This one-way region can be tuned based on the requirements by varying the magnetic bias or the temperature of operation. Similarly, if we change the direction of magnetic bias the light in the opposite direction gets blocked. This phenomenon can be exploited to realize switches at the terahertz frequencies.

\subsection{Magnonic modes}

\begin{figure}
        
   \centering
    \includegraphics[scale=0.36]{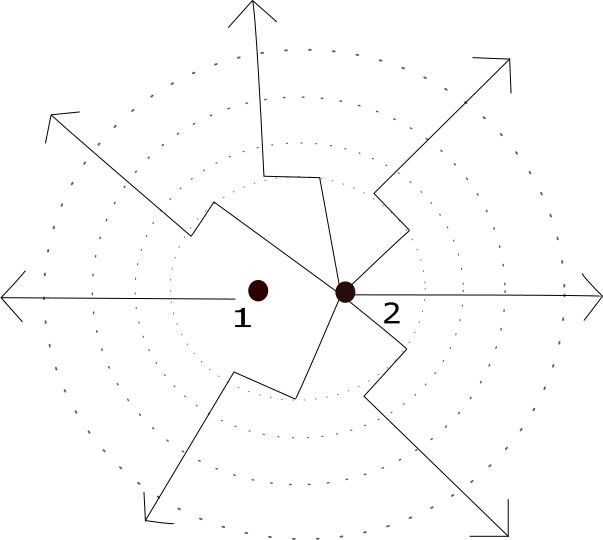}
    \caption{The Kink Model of Electromagnetic Radiation}
    \label{kink}
    
\end{figure}

Let us now look the same model from the perspective of radiating structures. The electromagnetic theory introduced by Maxwell in 1862 forms the basis of our understanding of the present day communication systems over a wide frequency spectrum of dc to optics. It is undeniable that it has led to rapid advancements in the field of electronics, radio waves and photonics over the course of time. The invention of antennas in 1880s enabled long distance communications which laid the foundations of the field of radiology. Today, fiber optics cables are used to transmit information at very high rates. It is also true that the onset of 21st century posed various limitations in the models being utilized to understand these devices \cite{1715229} \cite{article1} \cite{6379735}. One such fundamental problem that has gained significant interest over last few decades is understanding the process of radiation which does not seem to be well understood \cite{631830} \cite{730536} \cite{6347960}. There has been a perception among the scientific community that these engineered antennas are the source of radiation produced by such systems. Here, we have provided some insights on the processes that are involved in the near field of a dipole antenna that leads to far field radiation.

It is known that there are three regions around the dipole classified according to the rate of change of fields in the region \cite{10.5555/1208379}. We are usually more concerned for the far fields because of the notion of antennas as the radiating devices. The study of near fields for the hertz dipole in past has revealed interesting features such as the presence of causal surfaces (the surfaces with equal and opposite power flow) around the dipole \cite{924604} \cite{inbook}. These surfaces have been called as the source of the radiation. In this paper, we have analysed the near fields of a half wavelength dipole antenna and a loop antenna in frequency domain using Method of Moments. The method of causal surfaces has been applied to the dipole antenna and the location of these surfaces from the dipole has been calculated. It is shown that these surfaces oscillates with time between two fixed points thus forming a partial standing wave in the near field region \cite{9252662}. The calculation of impedance suggests its strong dependence with the distance from the antenna. It has been found that the near field region is divided into alternate regions of capacitive and inductive energy whose boundaries are determined by the location of causal surfaces. This region around the antenna is formed as soon as the current starts flowing through the antenna. This seat of reactive energy is the actual source of radiation \cite{article2} \cite{article3}.

The Kink Model of electromagnetic radiation as suggested by \cite{200289} can be employed to dive deeper in the near field region of a dipole antenna excited by an electric or magnetic feed. The electrons move back and forth in these devices which produces time varying current. We have taken these elements to compute the electric and magnetic fields around a dipole in Eq.\ref{9} and loop antenna in Eq.88. Suppose a charge is accelerated from point 2 to 1 as shown in Fig.\ref{kink}. This information will takes some time to reach 1. The initial E-field lines at 2 now points at 1. This would make the electric field lines discontinuous. A non-radial transverse component is needed to make the field lines continuous. This forms a wave-propagation front as shown in the figure. This is the radiation component produced by charge’s acceleration. The radial component of the electric field which falls exponentially with distance gives the reactive energy which can be electrostatic or inductive. This small seat of reactive region around an antenna which exists for distances upto few wavelengths for a half wavelength dipole is the near field region and acts as an imaginary radiating source for an antenna.

In the near field region of a hertzian dipole, the transverse component of the electric and magnetic fields are in phase quadrature with each other giving a net zero radial power flow. The electric field lines takes some time to completely orient along the transverse direction. The radial components of the electric field that exists in this region gives the power density in the transverse direction which accumulates the imaginary energy that exists in this region. The radial component of the electric field decays exponentially and ceases to exist beyond the near field limit. This is also the point when phase difference between the transverse field components $E_\theta$ and $H_\phi$ becomes zero.

\begin{figure}
     \centering   
    \includegraphics[scale=0.4]{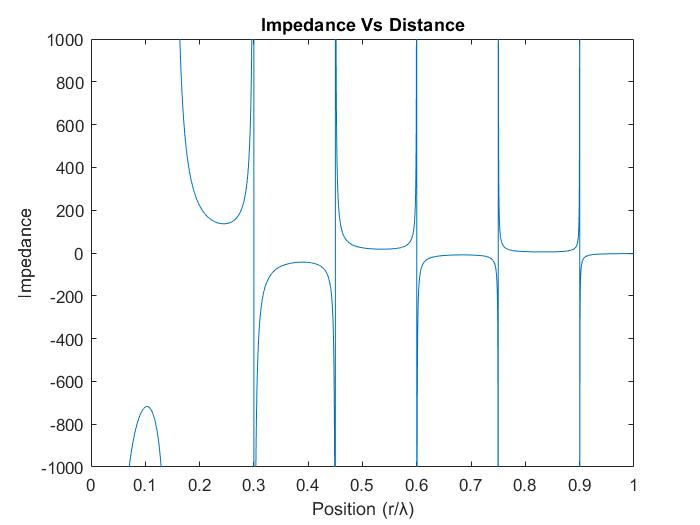}
    \caption{Wave impedance in the near field of a dipole antenna }
    \label{Impedance}
    
\end{figure}

The plot of wave impedance (the ratio of $E_\theta$ and $H_\phi$) in the near field region of an antenna as shown in Fig.\ref{Impedance} shows how this region is divided into small alternate sections containing electrostatic and inductive energy.  Whenever $H_\phi$ (which depends on distance) becomes zero, the impedance becomes infinite. These surfaces which separates these regions are the causal surfaces. The impedace in these region has been tabulated in Table 1. This can be clearly seen that the impedance in near field is the strong function of range from the antenna. This looks very similar to the transmission line as if such line is connected between the antenna and the far-field and the energy gets transmitted to the far-field through these lines. As we go away from the antenna, the impedance becomes constant equal to that of a free space.

\begin{figure}

\subfloat{\includegraphics[width = 3.5in]{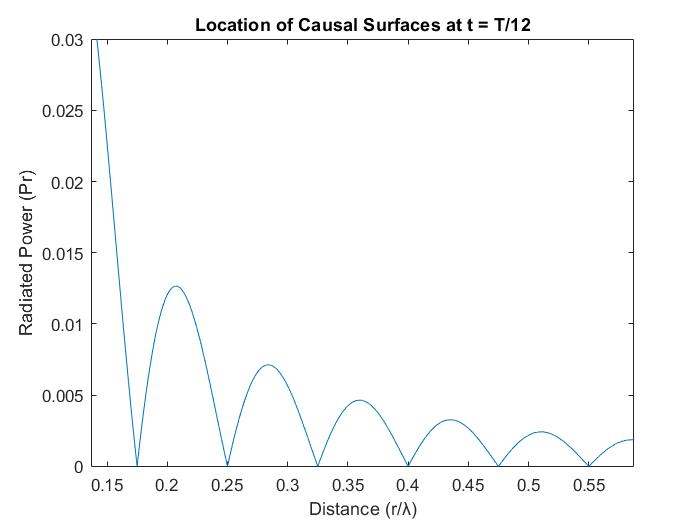}} 
\subfloat{\includegraphics[width = 3.5in]{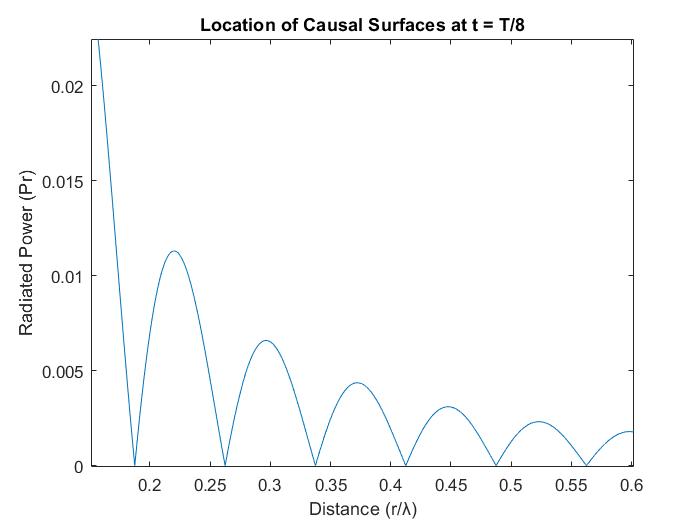}}\\
\subfloat{\includegraphics[width = 3.5in]{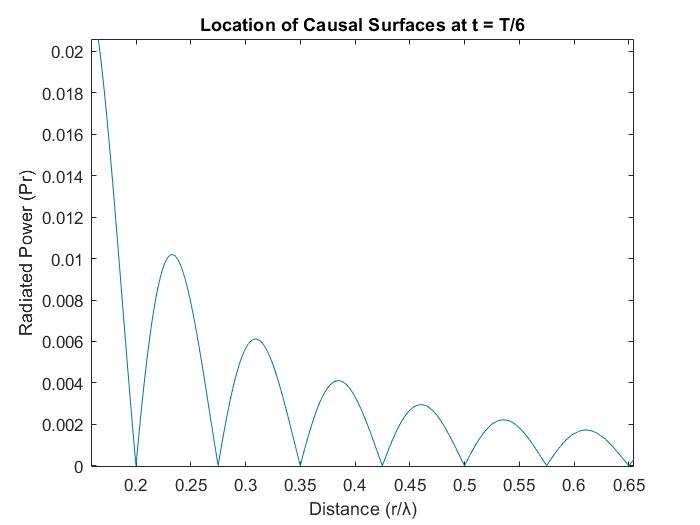}}
\subfloat{\includegraphics[width = 3.5in]{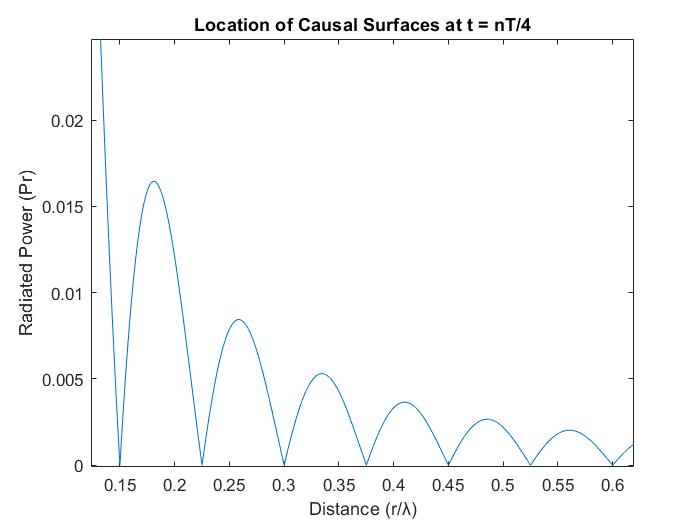}} 
\caption{Location of Causal Surfaces at different time instants}
\label{some example}
\end{figure}

Fig. \ref{some example} shows the plot for the location of causal surfaces at different time instants. It can be observed that these surfaces keeps moving to the next nodal point as the time progresses and takes a quarter of a period to completely move to the next location. In the next quarter, these surfaces again starts moving but backward to its original location. This process continues to happen two times in each time period. An electromagnetic wave is radiated to far field as soon as these surfaces completes one round trip. This phenomenon suggests that the nodal points (i.e the causal surfaces) keeps oscillating between two points during the process of radiation and thus forms partial standing waves in the near field region. We have also checked that the location of causal surfaces for a loop antenna and a dipole antenna are exactly same which confirms duality between these structures and further supports our results.

\begin{table}
  \begin{center}
    
    \label{tab:table1}
    \begin{tabular}{l|c|r} 
      \textbf{n} & \textbf{distance(d)} & \textbf{Impedance (r$<$d)}\\
      
      \hline
      1   & 0.075$\lambda$     & Capacitive\\
 2&   0.15$\lambda$  & Capacitive\\
 3 &0.225$\lambda$L & Inductive\\
 4    &0.3$\lambda$ & Inductive\\
 5&   0.375$\lambda$  & Capacitive\\
    \end{tabular}
    \caption{Electromagnetic energy around a dipole antenna}
    
  \end{center}
  \label{ccc}
\end{table}

We can now also look at the problem with a more rigorous approach. Let us take a excited dipole antenna. We know that these structure has oppositely charged metallic rods that leads to electric field lines between them. Let us take a positively charged particle and a negatively charged particle sitting at center of dipole. As soon as we excite the dipole, these particle will race towards the opposite ends and so will the field lines associated between them which later forms a radiated plane wave. There are many such particles in the antenna that contributes to the radiated energy and due to symmetry it seems that the radiated power should be directly proportional to the length of antenna but it varies as log of length for dipoles of larger length with uneven distribution of radiated power along the length of antenna \cite{1715229} \cite{730536}. Also, it seems that these field lines shouldn't have distinct locations but should be divided evenly across the near field region. But we have found the existence of causal surfaces around the antenna which oscillates between distinct locations around the antenna. These oscillations are formed due to the coupling of light with the magnetic energy of the medium which is supplied by the dipole in half of the period and then extracted in form of electrostatic current that flows in the dipole. This is the state where spin of the light gets locked to that of a excited state of electrons so called magnon and the oscillation of such state leads to radiation.

\section{Methods}

\subsection{Gyroelectric Waveguide}

In this section, we will deduce the permittivity tensor for a gyroelectric medium biased with homogeneous magnetic field as discussed in \cite{Caltech}. When a plane electromagnetic wave having electric field vector E and magnetic field vector B travels through such a medium, each electron having charge q and velocity v is subjected to a force which is given by the well known Lorentz equation:

\begin{equation}
    F = qE + q(v\times B_o)
    \label{lorentz}
\end{equation}

where $\mu_o$ is the permeability of free space. Applying Newton's law to each electron of mass m in \eqref{lorentz}, we get:

\begin{equation}
    m \frac{d\vec{v}}{dt} = q \vec{E} + q \vec{v} \times \vec{B_o}
    \label{force}
\end{equation}

In this equation, both E and v are functions of space and time, and $B_o$ is spatially uniform and independent of time. Here, we need to solve \eqref{force} for the velocity v. If we differentiate \eqref{force} with respect to time, we get:

\begin{equation}
   m \frac{d^2 \vec{v}}{dt^2} = q \frac{d \vec{E}}{dt} + q \frac{d}{dt} \left( \vec{v} \times \vec{B_o} \right)
\end{equation}

Now, if we multiply the equation with $\frac{q}{m^2}B_o$, and transform the right side using well-known vector identity, we get:

\begin{equation}
    \frac{q}{m}\dv[2]{(v\times B_o)}{t} = \frac{q^2}{m^2} \dv{(E\times B_o)}{t} + \frac{q^2}{m^2} \dv{(v\cdot B_o)}{t} B_o -  \frac{q^2}{m^2} (B_o \cdot B_o) \dv{v}{t}
    \label{eqqq}
\end{equation}

Multiplying \eqref{force} by $\frac{q^2}{m^3}B_o$, we get:

\begin{equation}
    \frac{q^2}{m^2}\dv{(v\cdot B_o)}{t}B_o = \frac{q^2}{m^3}(E\cdot B_o)B_o 
    \label{eqqqq}
\end{equation}

And operating on \eqref{force} with $\frac{1}{m}\pdv[2]{}{t}$ we get:

\begin{equation}
    \dv[3]{v}{t} = \frac{q}{m} \dv[2]{E}{t} + \frac{q}{m}\dv[2]{v \times B_o}{t}
    \label{eqqqqq}
\end{equation}

Using (28), (29) and (30), we get:

\begin{equation}
    \dv[3]{v}{t} + \left(\frac{q B_o}{m}\right)^2 \dv{v}{t} = \frac{q}{m} \dv[2]{E}{t} + \frac{q^2}{m^2}\dv{(E \times B_o)}{t} + \frac{q^3}{m^3}(E \cdot B_o)B_o
    \label{eqqqqqq}
\end{equation}

Since no term in (31) contains a product of time-dependent functions, we are free to restrict the time dependence to $e^{-i\omega t}$ by replacing $\pdv{}{t}$ by $-i\omega$, etc. Thus (31) becomes:

\begin{equation}
    v(i\omega^3 - i\omega(\frac{qB_o}{m})^2) = -\omega^2\frac{q E}{m} - i\omega \frac{q^2}{m^2} E\times B_o + \frac{q^3}{m^3}(E\cdot B_o)B_o 
    \label{eqqqqqqqq}
\end{equation}

Here, the term $(\frac{q}{m}B_o)^2$ is equal to the square of the gyrofrequency $\omega_g$. Now, we can introduce a rectangular coordinate system such that $B_o$ lies along the z-axis. If we denote the unit vectors along the coordinate axis by $u_x$, $u_y$, $u_z$, we can write $B_o$ = $a_z B_o$, $E \cdot B_o$ = $E_z B_o$, and $E \times B_o = a_x E_y B_o - a_y E_x B_o$. If we define the plasma frequency $\omega_o$ by $\frac{Nq^2}{m\epsilon_o}$, where N is the number of electrons per unit volume, we can write (32) in the following form:

\begin{equation}
    Nqv = -i\omega (\frac{-\epsilon_o \omega_o^2}{\omega^2 - \omega_g^2} E + \frac{\epsilon_o \omega_o^2 \omega_g}{i\omega(\omega^2 - \omega_g^2)} (a_z E_y - a_y E_x)  + \frac{\epsilon_o \omega_o^2 \omega_g^2}{\omega^2(\omega^2 - \omega_g^2)} a_z E_z)
    \label{eqqqqqqqqqq}
\end{equation}

In (33) Nqv is the convection current and we must add the displacement current -$i\omega \epsilon_o E$ in order to obtain the total current J:

\begin{equation}
   J = -i\omega \epsilon_o E + Nqv
    \label{eqqqqqqqqqq1}
\end{equation}

The x,y and z components of J are easily obtained from (33) and (34). They are

\begin{equation}
    J_x = -i\omega[\epsilon_o(1-\frac{\omega_o^2}{\omega^2 - \omega_g^2})E_x - i\epsilon_o \frac{\omega_o^2\omega_g}{\omega(\omega^2 - \omega_g^2)} E_y]
    \label{eqqqqqqqqqq2}
\end{equation}

\begin{equation}
    J_y = -i\omega[i\epsilon_o \frac{\omega_o^2\omega_g}{\omega(\omega^2 - \omega_g^2)}E_x + \epsilon_o(1-\frac{\omega_o^2}{\omega^2 - \omega_g^2}) E_y]
    \label{eqqqqqqqqqq3}
\end{equation}

\begin{equation}
    J_z = -i\omega[\epsilon_o(1-\frac{\omega_o^2}{\omega^2}) E_z]
    \label{eqqqqqqqqqq4}
\end{equation}

Equations (35), (36) and (37) can be written as:

\begin{equation}
    J_x = -i\omega\epsilon_{xx}E_x - i\omega\epsilon_{xy}E_y
    \label{eqqqqqqqqqq5}
\end{equation}

\begin{equation}
    J_y = -i\omega\epsilon_{yx}E_x - i\omega\epsilon_{yy}E_y
    \label{eqqqqqqqqqq6}
\end{equation}

\begin{equation}
    J_z = -i\omega\epsilon_{zz}E_z
\end{equation}

Now, Equations (38), (39) and (40) can be expressed as a tensor equation of the form:

\begin{equation}
    J = -i\omega \overset{\LARGE\frown}{\small{\epsilon}}\cdot E
\end{equation}

where $\overset{\LARGE\frown}{\small{\epsilon}}$ is the permittivity tensor of the form:

\begin{equation}
    \label{PermittivityTensork}
   \overset{\LARGE\frown}{\small{\epsilon}} = \left\lceil
\begin{matrix}
\epsilon_{xx} & -i\epsilon_{xy'} & 0\\
i\epsilon_{yx'} & \epsilon_{yy} & 0\\
0 & 0 & \epsilon_{zz}\\

\end{matrix}
\right\rceil
\end{equation}

where

\begin{equation}
   \epsilon_{xy'} = \epsilon_o \frac{\omega_o^2\omega_g}{\omega(\omega^2 - \omega_g^2)} = \epsilon_{yx'}
\end{equation}

\begin{equation}
   \epsilon_{xx} = \epsilon_o(1 -  \frac{\omega_o^2}{\omega^2 - \omega_g^2}) = \epsilon_{yy}
\end{equation}

\begin{equation}
   \epsilon_{zz} = \epsilon_o(1 -  \frac{\omega_o^2}{\omega^2}) 
\end{equation}

We can observe than whenever magnetic bias $B_o$ becomes zero, both the gyrotropic terms $\epsilon_{xy'}$ and $\epsilon_{xy'}$ also becomes equal to zero and the diagonal terms becomes equal to $\epsilon_o(1-\frac{\omega_o^2}{\omega^2})$. The medium is isotropic in this case.

The drude model for a free electron gas as discussed in the last section can be directly applied to the gyroelectric medium \cite{book}\cite{photonics8040133}\cite{article3}. In the presence of a uniform magnetostatic field, gyrotropy is introduced in the media resulting in the off-diagonal terms in the relative permittivity tensor. The permittivity tensor of such a medium can be written as:

\begin{equation}
    \label{PermittivityTensor}
   \overset{\LARGE\frown}{\small{\epsilon}} = \left\lceil
\begin{matrix}
\epsilon_1 & -j\kappa & 0\\
j\kappa & \epsilon_1 & 0\\
0 & 0 & \epsilon_2\\

\end{matrix}
\right\rceil
\end{equation}

here,

\begin{equation}
    \label{esilon1}
    \epsilon_1 = \epsilon_{\infty}(1 - \frac{\omega_p^2}{\omega^2 - \omega_c^2})
\end{equation}

\begin{equation}
    \label{esilon2}
    \epsilon_2 = \epsilon_{\infty}(1 - \frac{\omega_p^2}{\omega^2})
\end{equation}

\begin{equation}
    \label{kappa}
    \kappa =  \epsilon_{\infty}(\frac{\omega_c\omega_p^2}{\omega(\omega^2 - \omega_c^2)} )
\end{equation}

\hspace{15mm}where, \hspace{15mm}$\omega_c = \frac{eB_o}{m}$, called as cyclotron frequency and

\hspace{41mm}$\omega_p = \sqrt{(\frac{N_o e^2}{m\epsilon_o})} $ , called as plasma frequency

here, N is the electron density, $\epsilon_o$ is the permittivity of vacuum, $\epsilon_{\infty}$  is the background permittivity, which depends on the properties of the bound electrons in the material; m is electron's effective mass. In general, it is reasonable to assume the relative permittivity of the
gyroelectric medium $\mu$ = 1 since ‘natural’ materials do not present significant deviations from $\mu$ = 1 in a much wider frequency range.

We assume the material to be magnetically isotropic with relative permeability tensor $\mu = \mu[I3]$, where [I3] is a 3 × 3 identity matrix, and $\mu$ is the relative permeability constant

We can write the Maxwell's curl equations as:

\begin{equation}
    \label{MaxwellEqn3}
   \nabla \times E = i\omega \mu_o \mu H
\end{equation}

\begin{equation}
    \label{MaxwellEqn4}
   \nabla \times H = -i\omega \epsilon_o \overset{\LARGE\frown}{\small{\epsilon}} E
\end{equation}\\

Let us re-write these two Maxwell's equation in k-domain. Let us take k along the principal direction. We find a k-tensor $\overset{\LARGE\frown}{\small{k}} $ , such that the curl operation for the electric and magnetic field can be replaced with a matrix multiplication with the $\overset{\LARGE\frown}{\small{k}} $ , i.e., $\nabla$ ×E~ and $\nabla$ × H~ is equivalent
to $\overset{\LARGE\frown}{\small{k}} $ · E~ and $\overset{\LARGE\frown}{\small{k}} $ · H~ , respectively. The generalised $\overset{\LARGE\frown}{\small{k}} $ tensor in polar form can be written as:

\begin{equation}
    \label{ktensor}
   \overset{\LARGE\frown}{\small{k}} =  \left\lceil
\begin{matrix}
0 & -k_rcos\theta & k_rsin\theta sin\phi\\
k_rcos\theta & 0 & -k_rsin\theta cos\phi\\
-k_rsin\theta sin\phi & k_rsin\theta cos\phi & 0\\

\end{matrix}
\right\rceil
\end{equation}\\

Using this definition, we can write the Maxwell's curl equations as:\\

\begin{equation}
    \label{MaxwellEqn41}
   \overset{\LARGE\frown}{\small{k}} \cdot E = \omega \mu_o \mu H
\end{equation}

\begin{equation}
    \label{MaxwellEqn42}
  \overset{\LARGE\frown}{\small{k}} \cdot H = -\omega \epsilon_o \overset{\LARGE\frown}{\small{\epsilon}} E
\end{equation}

Substituting (54) in (53), we get:\\

\begin{equation}
    \label{CharEqn}
  [\overset{\LARGE\frown}{\small{k}} \cdot \overset{\LARGE\frown}{\small{k}} + k_o^2\mu \overset{\LARGE\frown}{\small{\epsilon}}]\cdot E = 0
\end{equation}

where $k_o = \omega \sqrt{\mu_0 \epsilon_0}$ is the free space wave vector, $\omega$ is the angular frequency, $\epsilon_o$ and $\mu_o$ are the absolute permittivity and permeability of free space, respectively.
$\overset{\LARGE\frown}{\small{k}} $ represents curl operation in the matrix form given by,

\begin{equation}
    \label{GeneralKMat}
  \overset{\LARGE\frown}{\small{k}} =  \left\lceil
\begin{matrix}
0 & -k_z & k_y\\
k_z & 0 & -k_x\\
-k_y & k_x & 0\\

\end{matrix}
\right\rceil
\end{equation}

In the above equation, \( k_x, k_y, \) and \( k_z \) represent the propagation constants along the \( x, y, \) and \( z \) directions, respectively. By equating the determinant of the matrix 
\[
\left[ \overset{\LARGE\frown}{\small{k}} \cdot \overset{\LARGE\frown}{\small{k}} + k_o^2 \mu \overset{\LARGE\frown}{\small{\epsilon}} \right]
\]
to zero, for a fixed frequency \( \omega \), we obtain the equation of isofrequency surfaces in polar form, as follows:

\begin{equation}
    \label{IsofrequencySurfacesEqn}
 \mu(\epsilon_2\mu(-\epsilon_1k_r^2k_o^4 + \epsilon_1^2k_o^6\mu - k_o^6\kappa^2\mu) + k_r^4k_o^2(\epsilon_1cos^2\theta + \epsilon_2sin^2\theta) -
 k_o^4k_r^2\mu((\epsilon_1^2 - \kappa^2)cos^2\theta + \epsilon_1\epsilon_2sin^2\theta)) = 0
\end{equation}

Solving this biquadratic equation, we get two independent iso-frequency surfaces given as:

\begin{equation}
    \begin{aligned}
        k_{r_1} &= 0.5 k_o \sqrt{\mu} \sqrt{ \frac{2 \epsilon_1 \epsilon_2 (1 + \sin^2 \theta) + (2 \epsilon_1^2 - 2 \kappa'^2) \cos^2 \theta}{\epsilon_1 \cos^2 \theta + \epsilon_2 \sin^2 \theta} } \\
        &\quad + 2 \sqrt{ \frac{-4 \epsilon_2 (\epsilon_1^2 - \kappa^2)(\epsilon_1 \cos^2 \theta + \epsilon_2 \sin^2 \theta) + \left( \epsilon_1 \epsilon_2 (1 + \sin^2 \theta) + (\epsilon_1^2 - \kappa^2) \cos^2 \theta \right)^2 }{ (\epsilon_1 \cos^2 \theta + \epsilon_2 \sin^2 \theta)^2 } }
    \end{aligned}
    \label{Surface1}
\end{equation}

\begin{equation}
    \begin{aligned}
        k_{r_2} &= 0.5 k_o \sqrt{\mu} \sqrt{ \frac{2 \epsilon_1 \epsilon_2 (1 + \sin^2 \theta) + (2 \epsilon_1^2 - 2 \kappa'^2) \cos^2 \theta}{\epsilon_1 \cos^2 \theta + \epsilon_2 \sin^2 \theta} } \\
        &\quad - 2 \sqrt{ \frac{-4 \epsilon_2 (\epsilon_1^2 - \kappa^2)(\epsilon_1 \cos^2 \theta + \epsilon_2 \sin^2 \theta) + \left( \epsilon_1 \epsilon_2 (1 + \sin^2 \theta) + (\epsilon_1^2 - \kappa^2) \cos^2 \theta \right)^2 }{ (\epsilon_1 \cos^2 \theta + \epsilon_2 \sin^2 \theta)^2 } }
    \end{aligned}
    \label{Surface2}
\end{equation}

The investigation of the nature of spin and gyrotropy imposed conditions in different topological regimes requires the computation of the spin along the isofrequency surfaces while restricting wave propagation to the \( x - y \) plane. Using Eq. (55), we can express the dependence of the \( y \) and \( z \) components of the electric field in terms of the \( x \)-component as follows:

\begin{equation}
    \label{CharEqnt}
 E_y = -\frac{i k_o^2 \kappa E_x}{\epsilon_1 k_o^2 - k_r^2}
\end{equation}

\begin{equation}
    \label{CharEqn2}
E_z = \frac{k_r^2 cos\theta sin\theta E_x}{k_r^2 sin^2\theta - \epsilon_2 k_o^2}
\end{equation}

Using (60) and (61), photonic spin along the isofrequency contour can be calculated using the Stokes polarization vector for electrical field as,

\begin{equation}
    \label{CharEqn3}
 Im(\bar{E} * E) = S_{ex}\hat{x} + S_{ey}\hat{x} + S_{ey}\hat{z}
\end{equation}

\begin{figure}
    \centering
    \includegraphics[scale = 0.4]{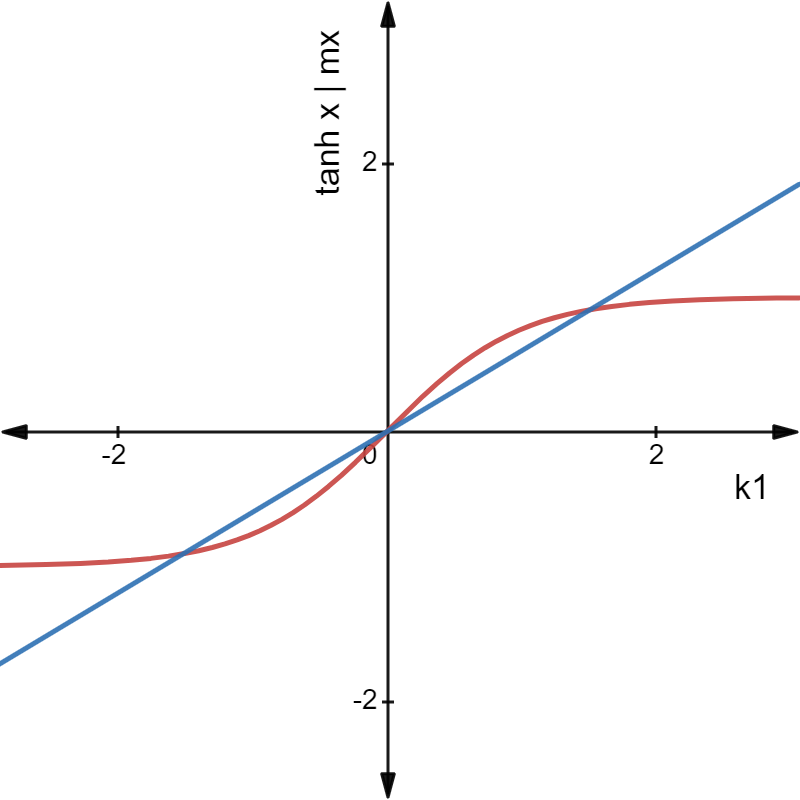}
    \caption{Plot of dispersion relation 1}
    \label{Plot of dispersion relation 1}
\end{figure}

Here,

\begin{equation}
\begin{aligned}
S_{ex} &= \frac{2\, {k_0}^2 \, \mathrm{kp} \, \mathrm{kr}^2 \, \cos(\theta) \, \sin(\theta) \, i}{ \left( \textrm{ep}_1 \, {k_0}^2 - \mathrm{kr}^2 \right) \, \sigma_1 \, \left( \frac{\left| k_0 \right|^4 \, \left| \mathrm{kp} \right|^2}{\left| \mathrm{kr}^2 - \textrm{ep}_1 \, {k_0}^2 \right|^2} + \frac{\left| \cos(\theta) \, \sin(\theta) \right|^2 \, \left| \mathrm{kr} \right|^4}{\left| \sigma_1 \right|^2} + 1 \right)} \\
\\
\textrm{where,} \\
\sigma_1 &= \mathrm{kr}^2 \, \sin^2(\theta) - \textrm{ep}_2 \, {k_0}^2 \\
\\
S_{ez} &= -\frac{2\, {k_0}^2 \, \mathrm{kp} \, i}{ \left( \textrm{ep}_1 \, {k_0}^2 - \mathrm{kr}^2 \right) \, \left( \frac{\left| k_0 \right|^4 \, \left| \mathrm{kp} \right|^2}{\left| \mathrm{kr}^2 - \textrm{ep}_1 \, {k_0}^2 \right|^2} + \frac{\left| \cos(\theta) \, \sin(\theta) \right|^2 \, \left| \mathrm{kr} \right|^4}{\left| \mathrm{kr}^2 \, \sin^2(\theta) - \textrm{ep}_2 \, {k_0}^2 \right|^2} + 1 \right)} \\
\\
S_{ey} &= 0
\end{aligned}
\end{equation}

Let us now take a gyroelectric core (I) sandwiched between two cladding layers ( II and III). If the two cladding layers are same, the dispersion relation for the surface plasmon polaritons at the two interfaces splits into equations namely,

\begin{figure}
    \centering
    \includegraphics[scale = 0.3]{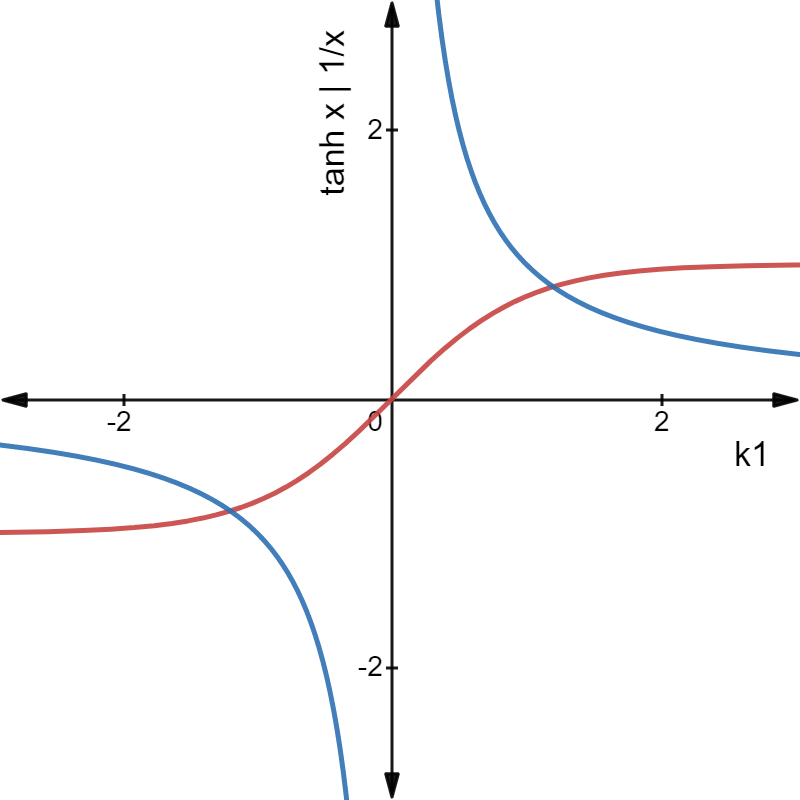}
    \caption{Plot of dispersion relation 2}
    \label{Plot of dispersion relation 2}
\end{figure}

\begin{equation}
    \label{disperson1}
    tanh (k1a) = \frac{-k1}{\epsilon}
\end{equation}

\begin{equation}
    \label{dispersion2}
    tanh (k1a) = \frac{-\epsilon}{k1}
\end{equation}

where, k1 is the wave vector in medium (I) along y direction, $\epsilon$ is the permittivity of medium (I) and a is the interface length. The solution of (\ref{disperson1}) and (\ref{dispersion2}) gives the oscillatory and the hyperbolic solutions respectively. In case of gyroelectric medium, the permittivity tensor becomes complex and the solution can also exist for complex wave vector k1. If we transform the permittivity tensor $\epsilon$ from real space to the hilbert's space using the Kramers-Kronig's relation, we get additional solutions for k1 in this space. The permittivity tensor for a gyroelectric equation can be split into two dependent tensors which are given by,

\begin{figure}
   \centering
    \includegraphics[scale = 0.3]{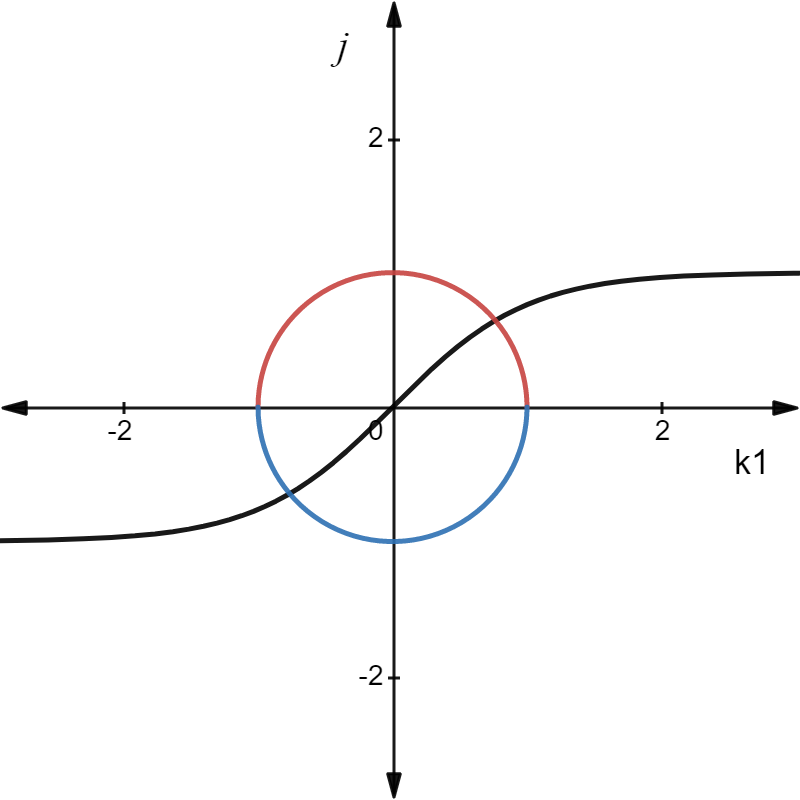}
    \caption{Plot of dispersion relation for complex k1}
    \label{Plot of dispersion relation for complex k1}
\end{figure}

\begin{equation}
    \label{PermittivityTensort}
  \Re(\epsilon) = \left\lceil
\begin{matrix}
\epsilon_x & 0 & 0\\
0 & \epsilon_y & 0\\
0 & 0 & \epsilon_z\\

\end{matrix}
\right\rceil
\end{equation}

\begin{equation}
    \label{PermittivityTensor1r}
  \Im(\epsilon) = \left\lceil
\begin{matrix}
0 & -\kappa & 0\\
\kappa & 0 & 0\\
0 & 0 & 0\\

\end{matrix}
\right\rceil
\end{equation}

Here, the imaginary part of gyroelectic tensor represents the spin tensor (S) associated in the medium which has two solutions of the form,\\

\begin{equation}
    \epsilon = \abs{\epsilon} e^{(+/-) j S} 
\end{equation}

 \begin{figure}
    \centering
    \includegraphics[scale = 0.6]{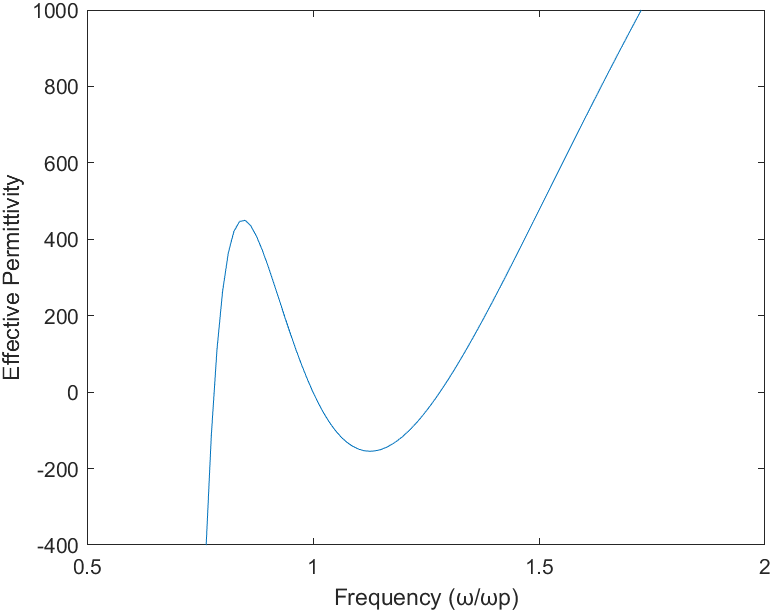}
    \caption{Plot of effective permittivity of the gyroelectric medium}
    \label{Plot of effective permittivity}
\end{figure}

\vspace{15mm}

The spin tensor can be split using the Pauli matrices into three dependent tensors characterizing the medium in hilbert's space which are given by,

\begin{equation}
    S_x = \left\lceil
\begin{matrix}
0 & 1\\
1 & 0\\

\end{matrix}
\right\rceil
\end{equation}

\begin{equation}
    S_y = \left\lceil
\begin{matrix}
0 & -j\\
j & 0\\

\end{matrix}
\right\rceil
\end{equation}

\begin{equation}
    S_z = \left\lceil
\begin{matrix}
1 & 0\\
0 & -1\\

\end{matrix}
\right\rceil
\end{equation}

The effective permittivity of the medium can now also be written as the sum of the real and imaginary tenors characterizing the medium in this space which is,

\begin{equation}
    \epsilon = (\epsilon_x \cross S_z) \Vec{y} + (\epsilon_y \cross S_z) \Vec{x} + j(\epsilon_z \cross S_x)
\end{equation}

This can also be written as:

\begin{equation}
    \epsilon = (\epsilon_x  + \epsilon_y + j\epsilon_z ) \cdot {(S_x + S_y + S_z)}
\end{equation}

Now, the effective permittivity of the medium can be written as the absolute value of the real part minus the imaginary spin associated in the medium. Summing the permittivity in all three directions gives the equation of the effective permittivity of the medium which is plotted in Fig.\ref{Plot of effective permittivity}. The figure clearly shows two different spin coupled states arising in the medium due to gyrotropy. Region 1 is the hyperbolic mode and Region 3 as shown in Fig.17 is the ellipsoidal mode in the medium whose topology depends on the anisotropy in the medium. Region2 leads to the evanescent modes where gyrotropy couldn't lead to any propagating solution. This is also the region of radiation as we will see later in the case of a dipole and a loop antenna. Region 4 is the optical frequency where we see two way propagation which is being exploited by the optical fibers. Now, the permittivity of the material can be written as,

\begin{equation}
    \label{PermittivityEq2}
  \abs{\epsilon} = (\epsilon_x)(\epsilon_y)(\epsilon_z) - \kappa^2
\end{equation}

This also gives the direct relationship between the wave vector k1 and the effective permittivity of the gyroelectric medium. Finally, we get,

\begin{equation}
    \label{dispersion3}
    \beta^2 = c(1-\frac{f(\omega)}{a}) + k0^2\epsilon(\omega)
\end{equation}

where, c = $\frac{1}{a^2}$ and k0 are constants, and f($\omega$) is a non-linear function of $\omega$, $\omega_c$ and $\omega_p$, $\omega_c$ is the cyclotron frequency and $\omega_p$ is the plasma frequency. Here, $\omega_c$ and $\omega_p$ can be controlled by varying magnetic bias and electron densities respectively. Now, let us assume the electric field vector $\psi(x,y)$ in the medium which is given by,

\begin{equation}
    \label{one}
    \psi(x,y) = Ae^{j(kx\cdot x + ky\cdot y)} + Be^{j(kx\cdot  x - ky\cdot  y)}
\end{equation}

Using (50) and (55), we get the following solutions for A $\&$ B,

\begin{equation}
    \label{twoeq}
    Ae^{j(kx\cdot x + ky\cdot y)} = Be^{j(kx\cdot  x - ky\cdot  y)}
\end{equation}

and
\begin{equation}
    \label{threer}
    A = \frac{\omega B_o \sqrt{\epsilon_x}}{2k0 \cdot \kappa}
\end{equation}

Substituting the values, we can see that A takes the undefined form of (0/0) in case we remove the magnetic bias from the medium. This suggests that such type of wave cannot be found be in normal isotropic medium and are topologically protected. We also get the equations describing the wave propagation vector in x and y directions, which are given by,

\begin{equation}
    \label{twoer}
    k_y = k0\sqrt{\epsilon_x}
\end{equation}

and
\begin{equation}
    \label{threee}
    k_x = \frac{-jk0 \cdot \kappa}{\sqrt{\epsilon_x}}
\end{equation}

Using the above equations, we can arrive at the equation of the wave propagating in the medium which is given by,

\begin{equation}
    \label{onee}
    \psi(x,y) = B_o\frac{\sqrt{\epsilon_1}}{\kappa}e^{j(kx\cdot x + ky\cdot y)}
\end{equation}

Now, let us define displacement vector for the electrical dipoles in the medium which is given by,

\begin{equation}
    \label{oneer}
    D = \epsilon_o(1 + \chi) E
\end{equation}

Here , the first term represents the permittivity of the material and the second term represents the induced permittivty due the magetic bias which has created a net dipole moment in the medium. Using semi-classical approximations where atoms are assumed to be quantized, we can also arrive at the similar solution for the wave propagating in such material. It should also be noted that the susceptibility $(\chi)$ has been derived using the damped oscillator model under the driving force which finally gives the Kramer's Kronig relations between the real and imaginary parts of the susceptibility tensor. We also get an additional relation between the permittivity tensor and the electric field in case we assume light to be also quantized which is quite intriguing. 

\begin{equation}
    \label{qcd}
   \epsilon \cdot E = 0
\end{equation}

It seems that the net charge bundle flowing in the medium has a spin associated with it which is getting conserved in the complex plane. The net charge in real plane is getting converted to spin in this space and the net spin remains conserved (i.e. the dipoles aren't permanently polarized). Now, let $[S_c]$ be the tensor relating charge with spin and let the charge bundle [J] be the column matrix. Then, we can get,

\begin{equation}
    \label{qcd2}
   \epsilon \cdot E = [S_c] \cdot [J]
\end{equation}

\subsection{Half wavelength dipole in frequency domain}

\begin{figure}
    \centering
    \includegraphics[scale=0.37]{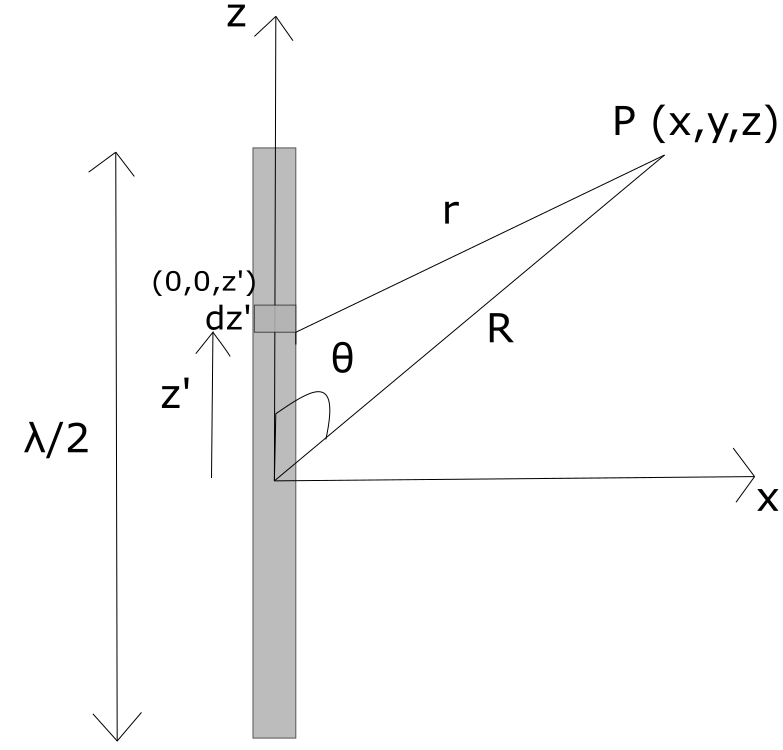}
    \caption{Half-wavelength dipole antenna in free space}
    \label{Half-wavelength_dipole}
\end{figure}

Let us assume a dipole antenna excited by a filamentary current source placed in a homogeneous source-free region. The current in the dipole as shown in Fig.\ref{Half-wavelength_dipole} is along the z axis, hence the magnetic vector potential also has to be along the z-axis. Let the unknown current be $I(z)$, which flows along the wire. The wire is assumed to be very thin. The z-component $E(z)$ of the radiated electric field at an arbitraty observation point \textbf{r}(z), is given by \cite{bookmom}.

\begin{equation}
    E_z(z) = \frac{j}{\omega \epsilon}[\frac{\partial^2 }{\partial z^2} + k^2]\int_{-L/2}^{L/2} I_z(z') \frac{e^{-jkr}}{4\pi r}dz'
\end{equation}

This is called Hallen's Integral equation. We may also move the differential operator under the integral sign, which gives the Pocklington's Integral equation:

\begin{equation}
    E_z(z) = \frac{j}{\omega \epsilon}\int_{-L/2}^{L/2}[\frac{\partial^2 }{\partial z^2} + k^2] I_z(z') \frac{e^{-jkr}}{4\pi r}dz'
\end{equation}

where, r is the distance between source and observation points, L is the length of the dipole, $\omega$ is the angular frequency, $\epsilon$ is the dielectric permittivity of the medium, and k is the wavenumber. Applying the boundary conditions for the electric field at a discrete set of N points (point matching) with discrete positions $z_m = \frac{-L}{2} + m\frac{L}{N}$, m = 1,...,N on the outer surface of the wire and approximating the current using step pulse basis function, we can model the above equation as a linear system of form:

\begin{figure}      
    \centering
    \includegraphics[scale = 0.4]{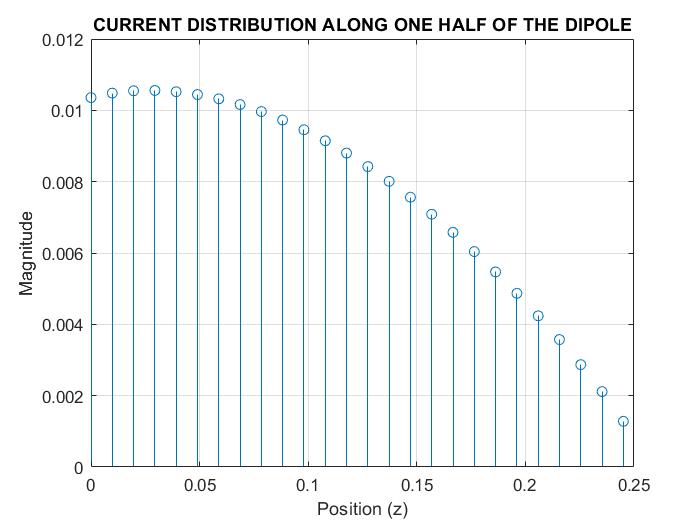}
    \caption{Current distribution along one half of the dipole calculated using MoM}
    \label{Current along dipole}   
\end{figure}

\begin{equation}
    \label{MOM2}
    [V] = [Z][I]
\end{equation}

where [I] is the vector of unknown weights for the current distribution on the wire which can be approximated as $I(z') = \sum_{n=1}^{n} I_iu_i$, where $u_i$ is a pulse function, $V_n = l$ and [Z] is a circulant matrix, namely [Z] = circ($z_1,z_2,...,z_n$) which is given by \cite{bookmom}. The solution of the current distribution for a half wavelength dipole is plotted as shown in Fig.\ref{Current along dipole}.

Now, we can write the near field equation for a hertz dipole as given in \cite{10.5555/1208379},

\begin{figure}
    \centering
    \includegraphics[scale = 0.4]{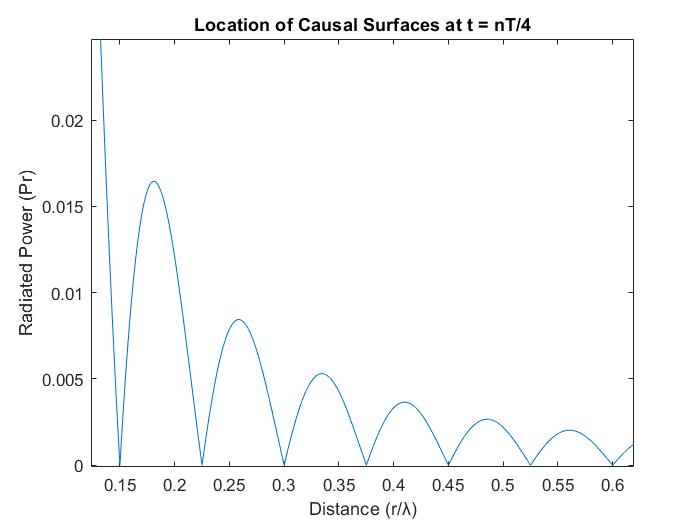}
   \caption{Location of Causal surfaces at t = nT/4}
    \label{causal_surfaces}
\end{figure}

\begin{equation}
    E_\theta = j\eta \frac{k I_o dl}{4\pi r} sin\theta [1 - \frac{j}{kr} -  \frac{1}{(kr)^2}] e^{-jkr}
\end{equation}

here $r=\sqrt{x^2 + y^2 + (z-z')^2}$. In spherical coordinates, the distance between two points P and the point on z-axis can be written as:

\begin{equation}
    |r2 - r1| = \sqrt{R^2 + z'^2 -2Rz'cos\theta}
\end{equation}

This implies, r = $\sqrt{R^2 + z'^2 -2Rz'cos\theta}$

We can assume that the finite antenna is made up of a large number of infinitesimal dipole. By the laws of superposition, we can sum the fields by the all such infintesimal dipole to find the field of a finite dipole antenna.

Let us take

\begin{equation}
    dE_\theta = j\eta \frac{k I_o dz'}{4\pi} sin\theta [\frac{1}{r} - \frac{j}{kr^2} -  \frac{1}{k^2r^3}] e^{-jkr}
\end{equation}

\begin{figure}
    \centering
    \includegraphics[scale = 0.4]{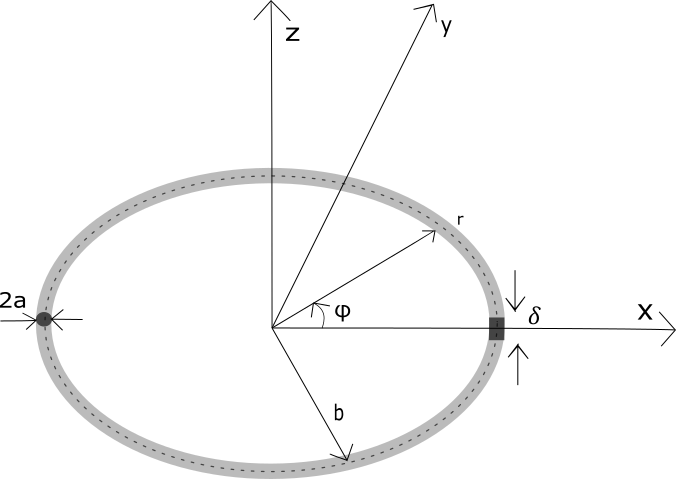}
   \caption{Thin Circular loop antenna}
    \label{fig:loop antenna}
\end{figure}

Now, for a $\lambda/2$ antenna, $E_\theta$ can be computed by summing all the small sections and using the above current distribution calculated using Method of Moments,

\begin{equation}
    E_\theta = j\eta \frac{k}{4\pi} sin\theta \int_{-\lambda/4}^{-\lambda/4} I_o(z')[\frac{1}{r} - \frac{j}{kr^2} -  \frac{1}{k^2r^3}] e^{-jkr} dz'
\end{equation}

Now, using discrete current distribution obtained using Method of Moments, $E_\theta$ can be calculated as,

\begin{equation}
    E_\theta = j\eta \frac{k}{4\pi} sin\theta \sum_{i=1}^{n} I(n)[\frac{1}{r} - \frac{j}{kr^2} -  \frac{1}{k^2r^3}] e^{-jkr} dz'
\end{equation}

here, n is number of segments. Similarly, $H_\phi$ can be calculated as:

\begin{equation}
    H_\phi = j\frac{k}{4\pi} sin\theta \sum_{i=1}^{n} I(n)[\frac{1}{r} - \frac{j}{kr^2}] e^{-jkr} dz'
    \label{9}
\end{equation}

We know that the radial power density, $W_r = E_\theta H_\phi$. This will be zero when either of $E_\theta$ or $H_\phi$ equals to zero. The surfaces (r) for which these are zero are called the causal surfaces. These surfaces doesn't have a net outward power flux and exists at the very vicinity of an antenna beyond which an electromagnetic wave is said to be radiated. In this paper \cite{924604}, Schantz has reported the existence of such surfaces and have analysed the near field region of a hertzian dipole in time domain. This region has been regarded as the seat of reactive energy which is the actual source of radiation.\\

\subsection{Loop antenna in frequency domain}

Lets assume a thin circular loop of radius b and wire radius a $<<$ b, driven by a delta gap generator as shown in Fig.\ref{fig:loop antenna}. Let the unknown current be $I(\phi)$, which flows along the wire axis. The wire is assumed to be very thin. The azimuthal component $E_{\phi}(r)$ of the radiated electric field at an arbitraty observation point \textbf{r}(r,$\phi$), is given by \cite{harrington1982field}.

\begin{multline}
    \label{$E_{phi}(r)$}
    E_{\phi}(r) = -j\omega\mu r' \int_{0}^{2\pi} I(\phi')cos(\phi-\phi')\frac{e^{-jk|r-r'|}}{4\pi|r-r'|} \,d\phi'\\
    +\frac{1}{j\omega\epsilon r}\int_{0}^{2\pi} \frac{dI(\phi')}{d\phi'}\frac{d}{d\phi'}\frac{e^{-jk|r-r'|}}{4\pi|r-r'|} \,d\phi'
\end{multline}

where, $|r-r'|$ is the distance between source and observation points, $\omega$ is the angular frequency, $\epsilon$ is the dielectric permittivity of the medium, $\mu$ is the magnetic permeability and k is the wavenumber. Applying the boundary conditions for the electric field at a discrete set of N points (point matching) with azimuthal positions $\phi_m = (\frac{2\pi}{N})m = m\phi$, m = 1,...,N on the outer surface of the wire and approximating the current using step pulse basis function, we can model the above equation as a linear system of form:

\begin{equation}
    \label{MOM3}
    [V] = [Z][I]
\end{equation}

where [I] is the vector of unknown weights, $V_n = (b+a)E_{\psi}(n\phi)$ and [Z] is a circulant matrix, namely [Z] = circ($z_1,z_2,...,z_n$) which is given by \cite{1603812}. Now, for a delta source of voltage V across the gap, located at $\phi = 0$,

\begin{equation}
    \label{Voltage}
    V_n = (1+\frac{a}{b})\frac{VN}{2\pi}\delta_{n,1}
\end{equation}

\begin{figure}
    \centering
    \includegraphics[scale = 0.4]{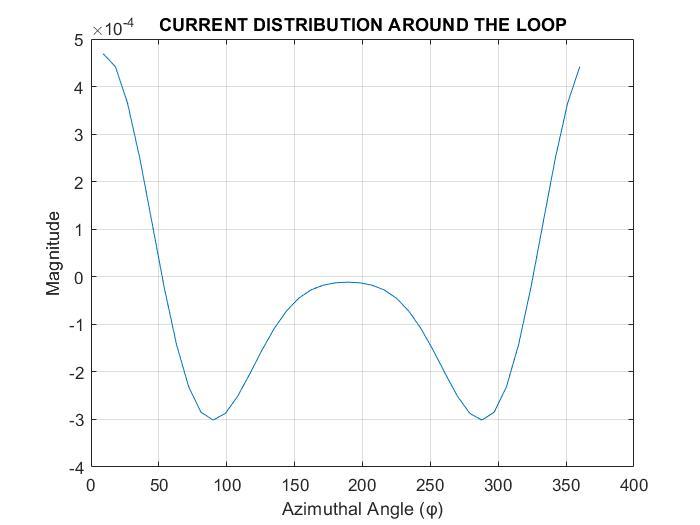}
   \caption{Current around the loop antenna calculated using MoM}
    \label{fig:currentloop antenna}
\end{figure}

where, $\delta_{n,1}$ is the kronecker delta function.

As described in \cite{1603812}, \eqref{MOM} can be solved analytically which gives the solution for current along the wire as:

\begin{equation}
    \label{Current}
    I_m = \frac{V}{2\pi}(1+\frac{a}{b})\sum_{q=1}^{N} \frac{1}{\lambda_q}e^{j(m-1)(q-1)\frac{2\pi}{N}}
\end{equation}

where,

\begin{equation}
    \label{EigenFunction}
    \lambda_q = \sum_{p=1}^{N} \omega^{(p-1)(q-1)}, \\
    q = 1,...,N, \omega = e^{\frac{j2\pi}{N}}
\end{equation}

The current around the loop is calculated using \eqref{Current} and plotted as shown in Fig.\ref{fig:currentloop antenna}. Now we can write the magnetic vector potential for an infintesimal element on a current carrying loop as given in \cite{10.5555/1208379},

\begin{equation}
    \label{A_phi}
    dA_\phi = \frac{\mu b I_m}{4\pi r} cos(\phi) e^{-jkr} d\phi
\end{equation}

here $r=\sqrt{x'^2 + y'^2 + z^2}$ for calculating fields on z-axis. In spherical coordinates, the distance between two points P on the antenna and the point on z-axis can be written as:

\begin{equation}
    \label{E_phi}
    r = |r2-r1| = \sqrt{b^2 + z^2}; 
\end{equation}

We can assume that the finite antenna is made up of a large number of infinitesimal antenna. By the laws of superposition, we can sum the fields by all such infintesimal antenna to find the field of a finite antenna. Using \eqref{A_phi}, we can get the magnetic field intensity for an infinitesimal element as,

\begin{equation}
   dH_\theta = - \frac{b k^2 I_o}{4\pi} [\frac{1}{r k^2 z} - \frac{z}{k^2r^3} + \frac{z}{j k r^2}] e^{-jkr} d\phi
\end{equation}

Now, $H_\theta$ can be computed at any point (0,0,z) by summing all the small sections and using the above current distribution calculated using Method of Moments,

\begin{equation}
    \label{E_theta summation}
    H_\theta =  -\frac{b k^2 }{4\pi}\int_{0}^{2\pi} I_{m}[\frac{1}{r k^2 z} - \frac{z}{k^2r^3} + \frac{z}{j k r^2}] e^{-jkr} d\phi
\end{equation}

Now, using discrete current distribution obtained using Method of Moments, $H_\theta$ can be calculated as:

\begin{equation}
    \label{E_theta summation2}
    H_\theta = -\frac{b k^2 }{2N}\sum_{m=1}^{N} I_{m}[\frac{1}{r k^2 z} - \frac{z}{k^2r^3} + \frac{z}{j k r^2}] e^{-jkr}
\end{equation}

\begin{figure}
    \centering
    \includegraphics[scale=0.36]{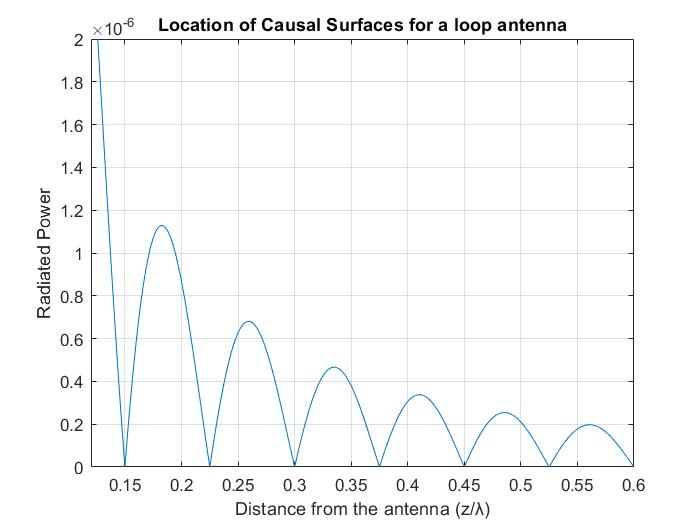}
    \caption{Location of causal surfaces for a loop antenna at t=nT/4}
    \label{causalloopantenna}
    
\end{figure}

here, N is the total number of segments. Similarly, $E_\phi$ can be calculated as:

\begin{equation}
    \label{H_phi summation2}
    E_\phi = \frac{\eta b k^2}{2 N} \sum_{m=1}^{N} I(m)[\frac{1}{r k^2 z} + \frac{z}{j k r^2}] e^{-jkr} 
\end{equation}

Here again, the radial power density for a loop antenna is calculated as, $W_r = -H_\theta E_\phi$. This will be zero when either of $E_\phi$ or $H_\theta$ equals to zero. The location of causal surfaces for a loop antenna is plotted as shown in Fig.\ref{causalloopantenna} .

\section{Results}
Here, we consider an indium antimony (InSb) with an external magnetic field as the gyroelectric medium. The material parameters used in the computation are $\mu_o$ = 1, $\epsilon_\infty$ = 15.68, $\omega_p$ = 8 THz, $\omega_c = 0.25\times\omega_p\times B$. These values are taken from \cite{axcv}. Here, B is the magnetic field. We have analyzed the resultant isofrequency contours and the photonic spin along the surface for two different values of B, 2T and 8T, one for which the cyclotron frequency is less than the plasma frequency ($\omega_c < \omega_p$) and the other for which the cyclotron frequency is more than the plasma frequency ($\omega_c > \omega_p$) . The Plot of $\epsilon1$, $\epsilon2$, and $\kappa$ for B = 2T is shown in Fig.\ref{fig:IsoContorsRegion} below. The plot also shows different regions defined by the terms in the permittivity tensor of the medium. We can observe that the characteristics of the propagating electromagnetic waves in such a medium depends on the relative magnitude of the isotropic and gyrotropic term. The photonic spin of the light is perfectly locked to that of a phonon in region 2 and 4. We can also observe the breaking of spin momentum locking in real space in regions 1 and 3 where the spin is getting locked in the complex space. This is the breathing mode so called beat modes and will be the major point of discussion in this section. It should be noted that in the absence of gyrotropy, the wave propagation in allowed only above the plasma frequency and the spin of the two isofrequency surfaces gets locked in the real plane. This the case of dielectric waveguides also used as optical fibers. There is no inherent handedness of the light in such a case which have arised in presence of a magnetic field.

\begin{figure}
\hspace{-55pt}
 \includegraphics[scale = 0.4]{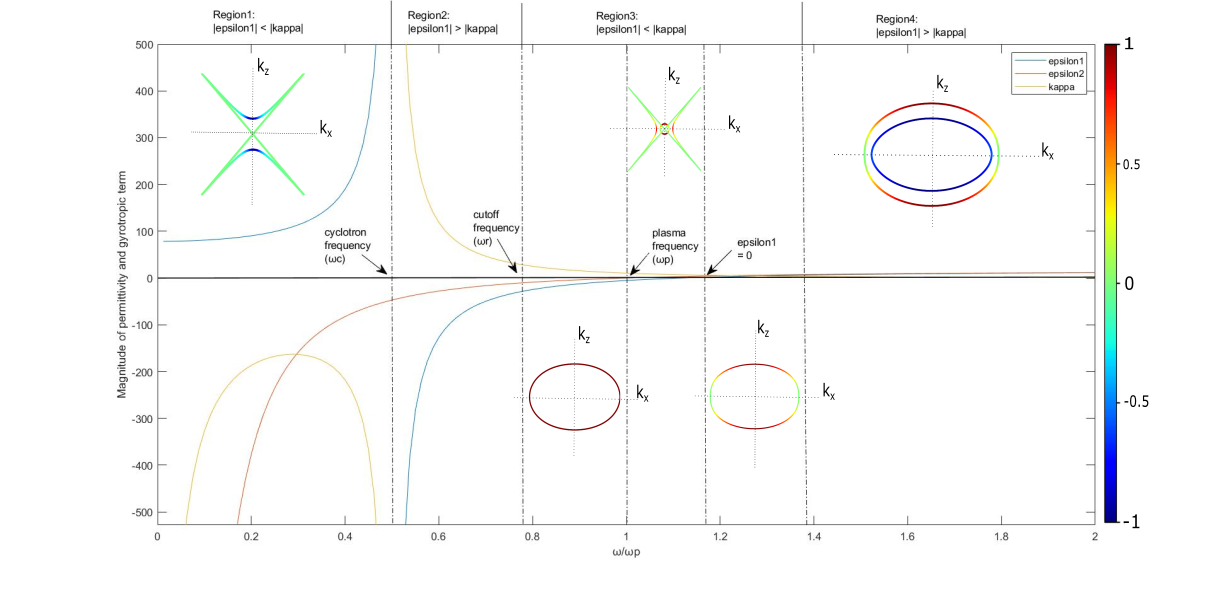}
\caption{The plot shows the dependence of $\epsilon1$(blue), $\epsilon2$(orange) and $\kappa$(yellow) with frequency. It also shows the resulting isofrequency contours and the corresponding photonic spin for the propagating electromagnetic waves at different frequency and the boundary separating these regions. The plot is generated for B=2T}
  \label{fig:IsoContorsRegion}
\end{figure}

It can be observed from Fig.\ref{fig:IsoContorsRegion} that the introduction of gyrotropy has led to the emergence of isofrequency surfaces well below the plasma frequency that doesn't exists without gyrotropy. The presence of magnetic bias also leads to suppression of the isofrequency surfaces in two of the regions. Table 2 shows different regions defined by the relative magnitude of the $\kappa$ and $\epsilon_1$. There are five frequencies of interest and each one of them can be exploited to design innovative non-reciprocal devices at THz frequencies. We will define each of the regions below and will show that there is a inherent polarisation of the light in such a medium which is locked to that of a phonon.

 \begin{table}
   \begin{center}

    \begin{tabular}{ |p{3cm}||p{3cm}||p{3cm}|}
 \hline
 Gyrotropic Term &$\epsilon_1 > 0$ & $\epsilon_1<0$\\
 \hline
 $\kappa = 0$   & One ellipsoid    &No solution\newline \\

 $|\epsilon_1| > |\kappa|$ &  Two concentric non-touching ellipsoids  & No solution \newline \\

 $|\epsilon_1| < |\kappa|$ & One hyperboloid/ One ellipsoidal & One hyperboloid and one ellipsoidal/ One ellipsoidal \\
 
 \hline
\end{tabular}
    
    \caption{Topological regimes for different values of $\epsilon_1$ as depicted in Fig.14}
    \label{tab:my_label1}
    \end{center} 
\end{table}

\begin{enumerate}
    \item Region 1: $\omega < \omega_c$\\
    
    In the absence of gyrotropy, there is no solution for the propagating electromagnetic waves in this region. The gyrotropy has led to the existence of real solution of the wave vector defined in Eq.58 and Eq.59. This can be explained by observing Eq.\ref{esilon1} and \ref{esilon2}. The introduction of cyclotron frequency makes $\epsilon_1$ positive in this region which would have been negative in absence of gyrotropy. It should be noted that the magnitude of gyrotropic term is greater than the anisotropic term in this region. The plot of isofrequency surface in this region can be observed from Fig.\ref{Iso1}.  The colormap of the contour represents the spin Sz, i.e., spin in the direction of the magnetic bias. It can be observed from the spin profile that the spin along z direction is anti-parallel to the material spin (+1). Here, the gyrotropy is large enough to suppress the mode which has a parallel spin-component. We can also see that the photonic spin along the isofrequency surface is symmetrical and is same for both forward and backward propagating waves. Thus, the propagating waves breaks spin-momentum locking.\\
    
\begin{figure}
    \begin{center}
        \includegraphics[scale = 0.4]{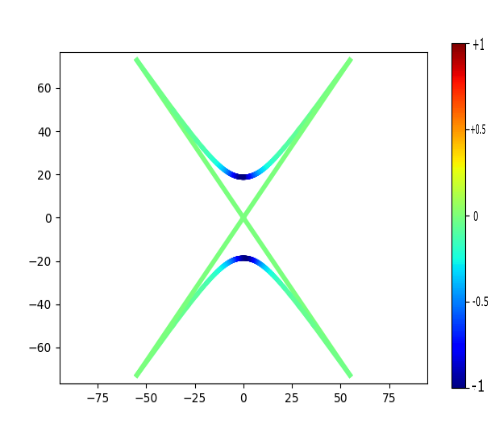}
  \caption{Spin Profile along the isofrequency contour for $\omega < \omega_c$. The color of the isofrequency contour represents the magnitude of z spin,
which is the direction of bias.}
\label{Iso1}
    \end{center}
 
  \label{fig:Isocontour1}
\end{figure}

    \item Region 2: $\omega_c < \omega < \omega_r$
    
    There is no real solution for the propagating wave between cyclotron frequency and the cutoff-frequency($\omega_r$) which is equal to $\frac{-\omega_c + \sqrt{\omega_c^2 + 4\omega_p^2}}{2}$. This is the cut-off frequency of the RHCP wave (having photonic spin +1) that exists above this frequency as can be seen from Fig.\ref{fig:IsoContorsRegion}. It should be noted that the magnitude of the diagonal term becomes larger that the gyrotropic term as we cross the cyclotron frequency. Here, the gyrotropy isn't enough to support any propagating mode.
    
    \item Region 3: $\omega_r < \omega < \omega_l$
    
    As we cross the right handed cutoff frequency ($\omega_r$), the gyrotropic term again becomes greater than the diagonal terms. We can observe that the diagonal terms are negative in this region which should not have allowed wave propagation in the region.  Here $\omega_l$ is called the left handed cutoff frequency for the LHCP wave (photonic spin -1) that exists beyond $\omega_l$ and is equal to $\frac{\omega_c + \sqrt{\omega_c^2 + 4\omega_p^2}}{2}$. This region is again divided into three sub-regions which is defined by the sign of diagonal terms $\epsilon1$ $\&$ $\epsilon2$. The photonic spin for each of the isofrequency contour is anti-parallel to the material induced spin (-1). We can clearly see the existence of gyrotropy induced spin waves in this region which wouldn't have existed without gyrotropy. The photonic spin across the RHCP wave (as can be seen from Fig.19) that exists in this region again breaks the spin-momentum locking in real space.
    
    \begin{figure}
    \begin{center}
        \includegraphics[scale = 0.4]{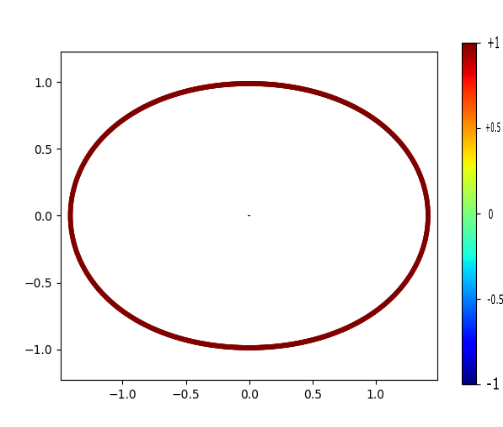}
  \caption{Spin Profile along the isofrequency contour for $\omega_r < \omega < \omega_l$. The color of the isofrequency contour represents the magnitude of z spin,
which is the direction of bias.}
    \end{center}
 
  \label{fig:Isocontour2rr}
\end{figure}

    There are two more frequencies of interest in this region. The plasma frequency ($\omega_p$) defines the frequency at which $\epsilon_2$ = 0. The frequency at which $\epsilon_1$ becomes zero is the epsilon zero frequency. As we cross the plasma frequency, we can see the existence of two isofrequency contours. Both has the photonic spin (+1) (as can be observed from Fig.20) . This is again anti-parallel to the material induced spin (-1). The resulting hyperbolic surface is because of the one of the diagonal term ($\epsilon2$) becoming positive in this region giving rise to the hyperbolic contour which has the different orientation than the other hyperbolic surface reported. This is because of the diagonal term along z direction is becoming positive this time resulting in the surface having spin parallel to the material induced spin along z direction which the gyrotropy couldn't suppress. The epsilon zero frequency is given by $\sqrt{\omega_c^2 + \omega_p^2}$. As we cross this frequency, we can observe the suppressing of the hyperbolic isofrequency surface in Fig.21 because of the diagonal elements along x and z direction becoming positive resulting in the material spin defined by only x and z direction which again breaks spin momentum locking in real space.
    
\begin{figure}
    \begin{center}
        \includegraphics[scale = 0.4]{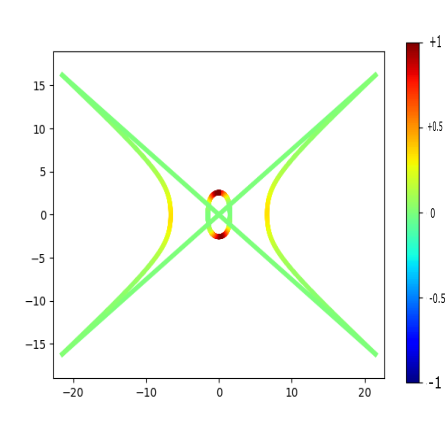}
  \caption{Spin Profile along the isofrequency contour for $\omega_r < \omega < \omega_l$. The color of the isofrequency contour represents the magnitude of z spin,
which is the direction of bias.}
    \end{center}
 
  \label{fig:Isocontour2x}
\end{figure}

\begin{figure}
\begin{center}
    \includegraphics[scale = 0.4]{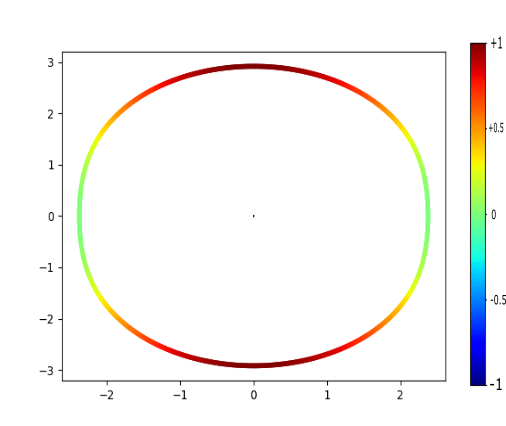}
  \caption{Spin Profile along the isofrequency contour for $\omega_r < \omega < \omega_l$. The color of the isofrequency contour represents the magnitude of z spin,
which is the direction of bias.}
\end{center}
 
  \label{fig:Isocontour2}
\end{figure}
    
    \item Region 4: $\omega_l < \omega$
    
    This is the frequency when the diagonal terms again becomes greater than the gyrotropic term. Two isofrequency surface exists in this region, having photonic spin (+1) and (-1) as can be observed from Fig.22. Since, the gyrotropic term is smaller here, it could not lead to suppression of any of the modes. Here, the spin of the two modes is perfectly locked in real space and is the perfect mode of operation for waveguiding in dielectric structures.
    
    \begin{figure}
    \begin{center}
        \includegraphics[scale = 0.4]{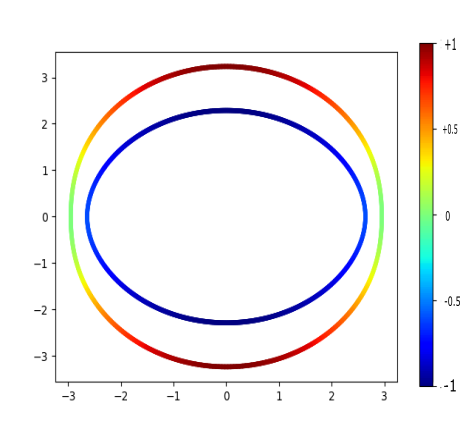}
  \caption{Spin Profile along the isofrequency contour for $\omega > \omega_l$. The color of the isofrequency contour represents the magnitude of z spin,
which is the direction of bias.}
    \end{center}
  \label{fig:Isocontour2r}
\end{figure}

\end{enumerate}

\begin{table}
   \begin{center}

    \begin{tabular}{ |p{3cm}||p{3cm}||p{3cm}|}
 \hline
 Gyrotropic Term &$\epsilon_1 > 0$ & $\epsilon_1<0$\\
 \hline
 $\kappa = 0$   & One ellipsoid    &No solution\newline \\

 $|\epsilon_1| > |\kappa|$ &  Two concentric non-touching ellipsoids/ One ellipsoid and one hyperboloid  & N/A \newline \\

 $|\epsilon_1| < |\kappa|$ & One hyperboloid/ One ellipsoidal & One hyperboloid and one ellipsoidal \\
 
 \hline
\end{tabular}
    
    \caption{Topological regimes for different values of $\epsilon_1$ as depicted in Fig. 20}
    \label{tab:my_label2}
    \end{center} 
\end{table}

Now, we will see how the increase in magnetic field changes the propagation characteristics of light in such a medium. We have taken the magnetic field equal to 8T. The cyclotron frequency is directly dependent on magnetic bias which shifts to a larger frequency if magnetic field is increased and can be larger than the plasma frequency. We can observe from Fig.23 that $\epsilon1$ is positive for frequencies less than cyclotron frequency in both the Region 1 and 2. The gyrotropic term is greater than  the anisotropic term in Region 1 which leads to the suppression of one of the propagating modes.

\begin{figure}
   \hspace{-15mm} \includegraphics[scale = 0.4]{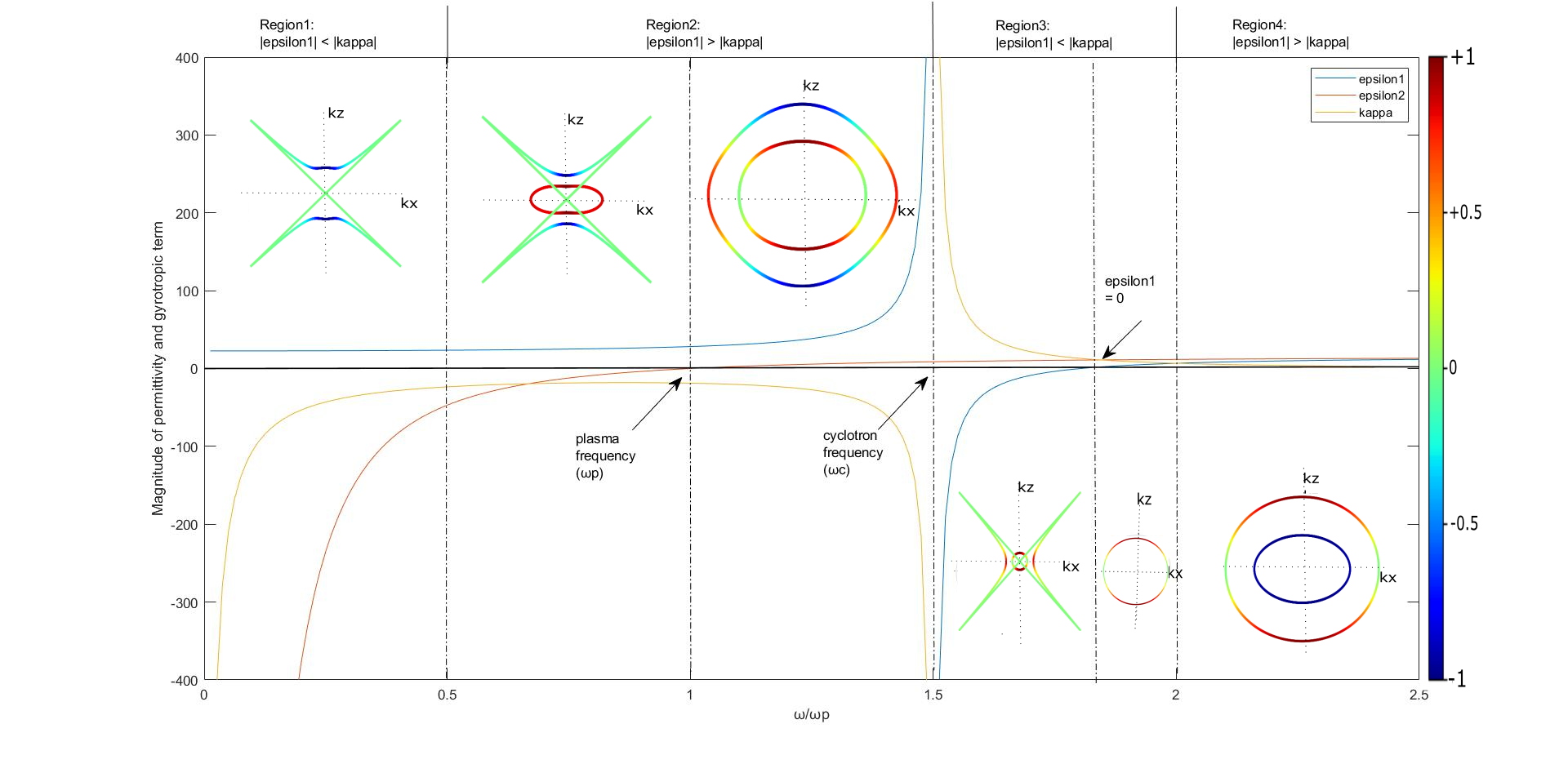}
  \caption{The plot shows the dependence of $\epsilon1$(blue), $\epsilon2$(orange) and $\kappa$(yellow) with frequency. It also shows the resulting isofrequency contours and the corresponding photonic spin for the propagating electromagnetic waves at different frequency and the boundary separating these regions. The plot is generated for B=8T}
  \label{fig:IsoContorsRegion2}
\end{figure}

As we cross the cyclotron frequency, we can see the existence of two isofrequency contours. Both has the photonic spin (+1) (as can be observed from Fig.23) . The resulting hyperbolic surface is because of the one of the diagonal term ($\epsilon2$) becoming positive in this region giving rise to the hyperbolic contour which has the different orientation than the other hyperbolic surface reported. This is because of the diagonal term along z direction is becoming positive this time resulting in the surface having spin parallel to the material induced spin along z direction which the gyrotropy couldn't suppress. The epsilon zero frequency is given by $\sqrt{\omega_c^2 + \omega_p^2}$. As we cross this frequency, we can observe the suppressing of the hyperbolic isofrequency surface because of the diagonal elements along x and z direction becoming positive resulting in the material spin defined by only x and z directions.

\begin{figure}
\begin{center}
    \includegraphics[scale=1]{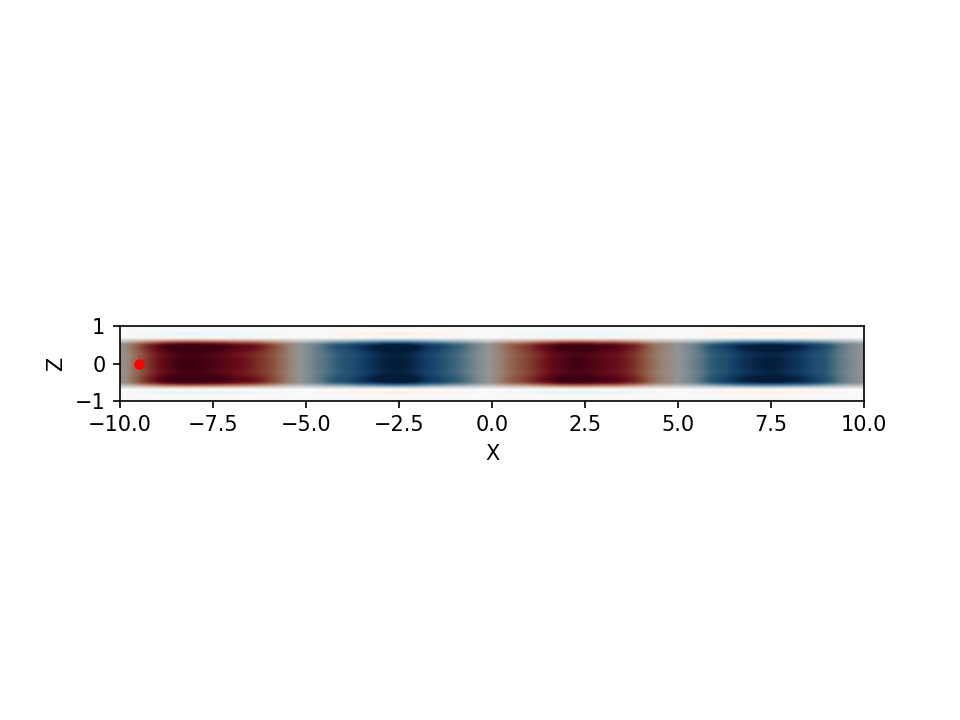}
  \caption{Gyroelectric Waveguide is simulated in MEEP at frequency of 10THz. Here, magnetic field is 4T.
}
\end{center}
  \label{fig:Gyrowaveguidesimulation}
\end{figure}
\section{Conclusions}

Our work concludes that gyrotropy leads to the realization of four major phenomenon associated with electromagnetic waves in a medium namely non-reciprocity, photonic spin, hyperbolic topology, and radiation which will play a key role in the design of the next generation photonic communication systems. Magnetically biased highly doped semiconductors such as InSb is a good candidate to realize non-reciprocal gyroelectric devices at THz frequencies. The drude model has been applied to get the permittivity tensor of the medium. The characteristics of propagating modes of light in such a medium has been determined using the Maxwell's equations. The propagation constant of the propagating wave has been plotted at different frequencies and it is shown how gyrotropy plays an important role in determining the existence of a particular isofrequency surface and the relevant modes of operation.

We calculated the photonic spin around the isofrequency contours and showed that these properties are inherent to the medium and are topologically protected. It is shown that the introduction of gyrotropy leads to the polarization of electrical dipoles of the medium in the direction that is locked to the magnetic bias. This interacts with the electromagnetic waves in the medium resulting into spin waves. These waves breaks the universal spin-momentum locking of enanescent waves in real space leading to the non-reciprocal character. It has also been shown that the relative magnitude of the diagonal and off-diagonal terms of the permittivity tensor determines the existence of such spin waves.

The design of a switch at THz frequency based on InSb(at 77K) is proposed. Here, gyroelectric medium is sandwiched between the three parallel dielectric waveguides. And we excite the cavity in the direction perpendicular to that of a light thus creating a phase difference between phonon and photon oscillations. Depending on the amplitude and phase  of these signals, the direction of magnetic bias, the electromagnetic wave that is launched in the middle waveguide can either be in the beat mode/breathing mode or the normal mode/decay mode. We have also simulated the breathing modes in a gyroelectric waveguide also called the beat mode using MEEP at the frequency of 10THz and magnetic bias of around 4T for case of rectangular waveguide with gyroelectric material in between which is shown in Fig.24.

In the case of an infinitesimal antenna, energy doesn’t propagate but gets transferred through magnon-photon oscillations to the near field which leads to radiation. The near field region deals with the evanescent waves which are confined to sub-wavelength distance and decays very rapidly. It contains imaginary power released by the antenna to the medium which oscillates between magnetic and electric form. The radiation fields are far reaching and propagating fields and carries real power. The impedance in the near field depends on the source and the distance from the antenna while impedance in the far field depends on the medium of propagation. The near field contains localized reactive energy oscillating at fixed points in space. The nodes ( position of zero power) oscillates between two fixed points and thus forms partial standing waves. These fixed points forms magnon-photon coupled states which results into the breathing modes. The energy flow velocity can be both smaller and larger than the velocity of light in the near field.

\bibliographystyle{unsrtnat}
\bibliography{references}  

\begin{thebibliography}{54}
\providecommand{\natexlab}[1]{#1}
\providecommand{\url}[1]{\texttt{#1}}
\expandafter\ifx\csname urlstyle\endcsname\relax
  \providecommand{\doi}[1]{doi: #1}\else
  \providecommand{\doi}{doi: \begingroup \urlstyle{rm}\Url}\fi

\bibitem[{Tonouchi}(2007)]{2007NaPho...1...97T}
Masayoshi {Tonouchi}.
\newblock {Cutting-edge terahertz technology}.
\newblock \emph{Nature Photonics}, 1\penalty0 (2):\penalty0 97--105, February 2007.
\newblock \doi{10.1038/nphoton.2007.3}.

\bibitem[Mittendorff et~al.(2021)Mittendorff, Winnerl, and Murphy]{Murphy}
Martin Mittendorff, Stephan Winnerl, and Thomas~E. Murphy.
\newblock 2d thz optoelectronics.
\newblock \emph{Advanced Optical Materials}, 9\penalty0 (3):\penalty0 2001500, 2021.
\newblock \doi{https://doi.org/10.1002/adom.202001500}.
\newblock URL \url{https://onlinelibrary.wiley.com/doi/abs/10.1002/adom.202001500}.

\bibitem[Chen et~al.(2021)Chen, Yu, and Gubbiotti]{Chen_2021}
Jilei Chen, Haiming Yu, and Gianluca Gubbiotti.
\newblock Unidirectional spin-wave propagation and devices.
\newblock \emph{Journal of Physics D: Applied Physics}, 55\penalty0 (12):\penalty0 123001, nov 2021.
\newblock \doi{10.1088/1361-6463/ac31f4}.
\newblock URL \url{https://doi.org/10.1088/1361-6463/ac31f4}.

\bibitem[Dai et~al.(2011)Dai, Xiang, Wen, and He]{xxxx}
Xiaoyu Dai, Yuanjiang Xiang, Shuangchun Wen, and Hongying He.
\newblock Thermally tunable and omnidirectional terahertz photonic bandgap in the one-dimensional photonic crystals containing semiconductor insb.
\newblock \emph{Journal of Applied Physics}, 109\penalty0 (5):\penalty0 053104, 2011.
\newblock \doi{10.1063/1.3549834}.

\bibitem[Maier et~al.(2001)Maier, Brongersma, Kik, Meltzer, Requicha, and Atwater]{article}
Stefan Maier, Mark Brongersma, Pieter Kik, Sheffer Meltzer, Ari Requicha, and Harry Atwater.
\newblock Plasmonics—a route to nanoscale optical devices.
\newblock \emph{Advanced Materials - ADVAN MATER}, 13, 10 2001.
\newblock \doi{10.1002/1521-4095(200110)13:193.0.CO;2-Z}.

\bibitem[Maier(2007)]{Maier}
SA~Maier.
\newblock Plasmonics: Fundamentals and applications.
\newblock page 245, 01 2007.

\bibitem[Ozbay(2006)]{Ozb}
Ekmel Ozbay.
\newblock Plasmonics: Merging photonics and electronics at nanoscale dimensions.
\newblock \emph{Science}, 311\penalty0 (5758):\penalty0 189--193, 2006.
\newblock \doi{10.1126/science.1114849}.

\bibitem[Schuller et~al.(2010)Schuller, Barnard, Cai, Jun, White, and Brongersma]{Schuller2010PlasmonicsFE}
Jon~A. Schuller, Edward~S. Barnard, Wenshan Cai, Young~Chul Jun, Justin~S. White, and Mark~L. Brongersma.
\newblock Plasmonics for extreme light concentration and manipulation.
\newblock \emph{Nature materials}, 9 3:\penalty0 193--204, 2010.

\bibitem[Monticone(2020)]{article5}
Francesco Monticone.
\newblock A truly one-way lane for surface plasmon polaritons.
\newblock \emph{Nature Photonics}, 14:\penalty0 461--465, 08 2020.
\newblock \doi{10.1038/s41566-020-0662-5}.

\bibitem[Eroglu(2010)]{book}
Abdullah Eroglu.
\newblock \emph{Wave Propagation and Radiation in Gyrotropic and Anisotropic Media}.
\newblock 01 2010.
\newblock ISBN 978-1-4419-6023-8.
\newblock \doi{10.1007/978-1-4419-6024-5}.

\bibitem[Sugimoto(1999)]{Sugimoto}
Mitsuo Sugimoto.
\newblock The past, present, and future of ferrites.
\newblock \emph{Journal of the American Ceramic Society}, 82\penalty0 (2):\penalty0 269--280, 1999.
\newblock \doi{https://doi.org/10.1111/j.1551-2916.1999.tb20058.x}.
\newblock URL \url{https://ceramics.onlinelibrary.wiley.com/doi/abs/10.1111/j.1551-2916.1999.tb20058.x}.

\bibitem[Lan et~al.(2015)Lan, Bi, Zhou, and Li]{article4}
Chu Lan, Ke~Bi, Ji~Zhou, and Bo~Li.
\newblock Experimental demonstration of hyperbolic property in conventional material—ferrite.
\newblock \emph{Applied Physics Letters}, 107:\penalty0 211112, 11 2015.
\newblock \doi{10.1063/1.4936612}.

\bibitem[Geiler and Harris(2014)]{6891475}
Anton Geiler and Vince Harris.
\newblock Atom magnetism: Ferrite circulators—past, present, and future.
\newblock \emph{IEEE Microwave Magazine}, 15\penalty0 (6):\penalty0 66--72, 2014.
\newblock \doi{10.1109/MMM.2014.2332411}.

\bibitem[Davis and Sloan(1993)]{276774}
L.E. Davis and R.~Sloan.
\newblock Semiconductor junction circulators.
\newblock In \emph{1993 IEEE MTT-S International Microwave Symposium Digest}, pages 483--486 vol.1, 1993.
\newblock \doi{10.1109/MWSYM.1993.276774}.

\bibitem[Yong et~al.(2001)Yong, Sloan, and Davis]{925497}
C.K. Yong, R.~Sloan, and L.E. Davis.
\newblock A ka-band indium-antimonide junction circulator.
\newblock \emph{IEEE Transactions on Microwave Theory and Techniques}, 49\penalty0 (6):\penalty0 1101--1106, 2001.
\newblock \doi{10.1109/22.925497}.

\bibitem[Ng et~al.(2004)Ng, Davis, and Sloan]{1266872}
Z.M. Ng, L.E. Davis, and R.~Sloan.
\newblock Measurements of v-band n-type insb junction circulators.
\newblock \emph{IEEE Transactions on Microwave Theory and Techniques}, 52\penalty0 (2):\penalty0 482--488, 2004.
\newblock \doi{10.1109/TMTT.2003.821924}.

\bibitem[Jawad and Sloan(2015)]{Jawad2015MillimetreWS}
G.~N. Jawad and R.~Sloan.
\newblock Millimetre wave semiconductor based isolators and circulators.
\newblock 2015.

\bibitem[Jawad et~al.(2017)Jawad, Duff, and Sloan]{7836295}
Ghassan~N. Jawad, Christopher~I. Duff, and Robin Sloan.
\newblock A millimeter-wave gyroelectric waveguide isolator.
\newblock \emph{IEEE Transactions on Microwave Theory and Techniques}, 65\penalty0 (4):\penalty0 1249--1256, 2017.
\newblock \doi{10.1109/TMTT.2016.2640298}.

\bibitem[Chow(1962)]{1137884}
Y.~Chow.
\newblock A note on radiation in a gyro-electric-magnetic medium--an extension of bunkin's calculation.
\newblock \emph{IRE Transactions on Antennas and Propagation}, 10\penalty0 (4):\penalty0 464--469, 1962.
\newblock \doi{10.1109/TAP.1962.1137884}.

\bibitem[Jawad et~al.(2015)Jawad, Sloan, and Missous]{7087406}
Ghassan~N. Jawad, Robin Sloan, and Mohamed Missous.
\newblock On the design of gyroelectric resonators and circulators using a magnetically biased 2-d electron gas (2-deg).
\newblock \emph{IEEE Transactions on Microwave Theory and Techniques}, 63\penalty0 (5):\penalty0 1512--1517, 2015.
\newblock \doi{10.1109/TMTT.2015.2418207}.

\bibitem[Shekhar et~al.(2019)Shekhar, Pendharker, Vick, Malac, and Jacob]{Shekhar:19}
Prashant Shekhar, Sarang Pendharker, Douglas Vick, Marek Malac, and Zubin Jacob.
\newblock Fast electrons interacting with a natural hyperbolic medium: bismuth telluride.
\newblock \emph{Opt. Express}, 27\penalty0 (5):\penalty0 6970--6975, Mar 2019.

\bibitem[Pendharker et~al.(2018)Pendharker, Kalhor, Mechelen, Jahani, Nazemifard, Thundat, and Jacob]{Pendharker:18}
Sarang Pendharker, Farid Kalhor, Todd~Van Mechelen, Saman Jahani, Neda Nazemifard, Thomas Thundat, and Zubin Jacob.
\newblock Spin photonic forces in non-reciprocal waveguides.
\newblock \emph{Opt. Express}, 26\penalty0 (18):\penalty0 23898--23910, Sep 2018.

\bibitem[Sen and Pendharker(2022)]{Sen_2022}
Rajarshi Sen and Sarang Pendharker.
\newblock Gyrotropy-governed isofrequency surfaces and photonic spin in gyromagnetic media.
\newblock \emph{Physical Review A}, 105\penalty0 (2), Feb 2022.

\bibitem[Wang and Wang(2019)]{articlex}
Neng Wang and Guo Wang.
\newblock Effective medium theory with closed-form expressions for bi-anisotropic optical metamaterials.
\newblock \emph{Optics Express}, 27:\penalty0 23739, 08 2019.
\newblock \doi{10.1364/OE.27.023739}.

\bibitem[Tuz et~al.(2017)Tuz, Fesenko, and Fedorin]{article1}
Vladimir Tuz, Volodymyr Fesenko, and Illia Fedorin.
\newblock Bi-hyperbolic isofrequency surface in a magnetic-semiconductor superlattice.
\newblock \emph{Optics Letters}, 42:\penalty0 4561, 11 2017.
\newblock \doi{10.1364/OL.42.004561}.

\bibitem[Joannopoulos et~al.(2011)Joannopoulos, Johnson, Winn, and Meade]{JoannopoulosJohnsonWinnMeade+2011}
John~D. Joannopoulos, Steven~G. Johnson, Joshua~N. Winn, and Robert~D. Meade.
\newblock \emph{Photonic Crystals: Molding the Flow of Light - Second Edition}.
\newblock Princeton University Press, 2011.
\newblock ISBN 9781400828241.
\newblock \doi{doi:10.1515/9781400828241}.
\newblock URL \url{https://doi.org/10.1515/9781400828241}.

\bibitem[Barnes et~al.(2003)Barnes, Dereux, and Ebbesen]{Barnes2003SurfacePS}
William~L. Barnes, Alain Dereux, and Thomas~W. Ebbesen.
\newblock Surface plasmon subwavelength optics.
\newblock \emph{Nature}, 424:\penalty0 824--830, 2003.

\bibitem[Hochberg et~al.(2004)Hochberg, Baehr-Jones, Walker, and Scherer]{Hochberg:04}
Michael Hochberg, Tom Baehr-Jones, Chris Walker, and Axel Scherer.
\newblock Integrated plasmon and dielectric waveguides.
\newblock \emph{Opt. Express}, 12\penalty0 (22):\penalty0 5481--5486, Nov 2004.
\newblock \doi{10.1364/OPEX.12.005481}.
\newblock URL \url{http://www.osapublishing.org/oe/abstract.cfm?URI=oe-12-22-5481}.

\bibitem[Haldane and Raghu(2008)]{articlehh}
F~Haldane and S~Raghu.
\newblock Possible realization of directional optical waveguides in photonic crystals with broken time-reversal symmetry.
\newblock \emph{Physical review letters}, 100:\penalty0 013904, 02 2008.
\newblock \doi{10.1103/PhysRevLett.100.013904}.

\bibitem[Ghosh and Pendharker(2022)]{Ghosh2022}
Niloy Ghosh and Sarang Pendharker.
\newblock {Electro-Optic beamforming in seamless wireless-to-photonic receiver arrays}.
\newblock 2 2022.

\bibitem[Wolz et~al.(2020)Wolz, Stehli, and Schneider]{nature}
T.~Wolz, A.~Stehli, and A.~Schneider.
\newblock Introducing coherent time control to cavity magnon-polariton modes.
\newblock \emph{Nature Physics}, 3\penalty0 (3):\penalty0 e266, 2020.
\newblock \doi{https://doi.org/10.1038/s42005-019-0266-x}.

\bibitem[{Wang} et~al.(2010){Wang}, {Belyanin}, {Crooker}, {Mittleman}, and {Kono}]{2010NatPh...6..126W}
X.~{Wang}, A.~A. {Belyanin}, S.~A. {Crooker}, D.~M. {Mittleman}, and J.~{Kono}.
\newblock {Interference-induced terahertz transparency in a semiconductor magneto-plasma}.
\newblock \emph{Nature Physics}, 6\penalty0 (2):\penalty0 126--130, February 2010.
\newblock \doi{10.1038/nphys1480}.

\bibitem[Brion et~al.(1972)Brion, Wallis, Hartstein, and Burstein]{PhysRevLett.28.1455}
J.~J. Brion, R.~F. Wallis, A.~Hartstein, and E.~Burstein.
\newblock Theory of surface magnetoplasmons in semiconductors.
\newblock \emph{Phys. Rev. Lett.}, 28:\penalty0 1455--1458, May 1972.
\newblock \doi{10.1103/PhysRevLett.28.1455}.
\newblock URL \url{https://link.aps.org/doi/10.1103/PhysRevLett.28.1455}.

\bibitem[Yu et~al.(2008)Yu, Veronis, Wang, and Fan]{PhysRevLett.100.023902}
Zongfu Yu, Georgios Veronis, Zheng Wang, and Shanhui Fan.
\newblock One-way electromagnetic waveguide formed at the interface between a plasmonic metal under a static magnetic field and a photonic crystal.
\newblock \emph{Phys. Rev. Lett.}, 100:\penalty0 023902, Jan 2008.
\newblock \doi{10.1103/PhysRevLett.100.023902}.
\newblock URL \url{https://link.aps.org/doi/10.1103/PhysRevLett.100.023902}.

\bibitem[Hu et~al.(2012)Hu, Wang, and Zhang]{Hu:12}
Bin Hu, Qi~Jie Wang, and Ying Zhang.
\newblock Broadly tunable one-way terahertz plasmonic waveguide based on nonreciprocal surface magneto plasmons.
\newblock \emph{Opt. Lett.}, 37\penalty0 (11):\penalty0 1895--1897, Jun 2012.
\newblock \doi{10.1364/OL.37.001895}.
\newblock URL \url{http://www.osapublishing.org/ol/abstract.cfm?URI=ol-37-11-1895}.

\bibitem[Fan et~al.(2012)Fan, Chang, Gu, Wang, and Chen]{6316071}
Fei Fan, Sheng-Jiang Chang, Wen-Hao Gu, Xiang-Hui Wang, and Ai-Qi Chen.
\newblock Magnetically tunable terahertz isolator based on structured semiconductor magneto plasmonics.
\newblock \emph{IEEE Photonics Technology Letters}, 24\penalty0 (22):\penalty0 2080--2083, 2012.
\newblock \doi{10.1109/LPT.2012.2219858}.

\bibitem[Miller(2006)]{1715229}
E.K. Miller.
\newblock Comparison of the radiation properfies of a sinusoidal current filament and a pec dipole of near-zero radius.
\newblock \emph{IEEE Antennas and Propagation Magazine}, 48\penalty0 (4):\penalty0 37--47, 2006.
\newblock \doi{10.1109/MAP.2006.1715229}.

\bibitem[Dumin et~al.(2012)Dumin, Volvach, and Dumina]{6379735}
O.~M. Dumin, I.~S. Volvach, and O.~O. Dumina.
\newblock Transient near field of hertzian dipole.
\newblock In \emph{2012 6th International Conference on Ultrawideband and Ultrashort Impulse Signals}, pages 69--71, 2012.
\newblock \doi{10.1109/UWBUSIS.2012.6379735}.

\bibitem[Miller(1997)]{631830}
E.K. Miller.
\newblock An exploration of radiation physics in electromagnetics.
\newblock In \emph{IEEE Antennas and Propagation Society International Symposium 1997. Digest}, volume~3, pages 2048--2051 vol.3, 1997.
\newblock \doi{10.1109/APS.1997.631830}.

\bibitem[Smith(1998)]{730536}
G.S. Smith.
\newblock On the interpretation for radiation from simple current distributions.
\newblock \emph{IEEE Antennas and Propagation Magazine}, 40\penalty0 (4):\penalty0 39--44, 1998.
\newblock \doi{10.1109/74.730536}.

\bibitem[Kaiser(2012)]{6347960}
Gerald Kaiser.
\newblock The reactive energy of transient em fields.
\newblock In \emph{Proceedings of the 2012 IEEE International Symposium on Antennas and Propagation}, pages 1--2, 2012.
\newblock \doi{10.1109/APS.2012.6347960}.

\bibitem[Balanis(2005)]{10.5555/1208379}
Constantine~A. Balanis.
\newblock \emph{Antenna Theory: Analysis and Design}.
\newblock Wiley-Interscience, USA, 2005.
\newblock ISBN 0471714623.

\bibitem[Schantz(2001)]{924604}
H.G. Schantz.
\newblock Electromagnetic energy around hertzian dipoles.
\newblock \emph{IEEE Antennas and Propagation Magazine}, 43\penalty0 (2):\penalty0 50--62, 2001.
\newblock \doi{10.1109/74.924604}.

\bibitem[Schantz et~al.(2002)Schantz, Smith, and Cloude]{inbook}
Hans Schantz, Paul Smith, and Shane Cloude.
\newblock \emph{On the Localization of Electromagnetic Energy}, pages 89--96.
\newblock 01 2002.
\newblock ISBN 0-306-47338-0.
\newblock \doi{10.1007/0-306-47948-6_11}.

\bibitem[Naydenko(2020)]{9252662}
Victor Naydenko.
\newblock Velocity of the fields radiated hertz dipole excited by gaussian pulse.
\newblock In \emph{2020 IEEE Ukrainian Microwave Week (UkrMW)}, pages 154--160, 2020.
\newblock \doi{10.1109/UkrMW49653.2020.9252662}.

\bibitem[Tang et~al.(2008)Tang, Kocabas, Latif, Okyay, Ly-Gagnon, Saraswat, and Miller]{article2}
Liang Tang, Sukru Kocabas, Salman Latif, Ali Okyay, Dany Ly-Gagnon, Krishna Saraswat, and David Miller.
\newblock Nanometre-scale germanium photodetector enhanced by a near-infrared dipole antenna.
\newblock \emph{Nature Photonics}, 2:\penalty0 226--229, 03 2008.
\newblock \doi{10.1038/nphoton.2008.30}.

\bibitem[Eroglu and Lee(2006)]{article3}
Abdullah Eroglu and Jay Lee.
\newblock Wave propagation and dispersion characteristics for a nonreciprocal electrically gyrotropic medium.
\newblock \emph{Progress in Electromagnetics Research-pier - PROG ELECTROMAGN RES}, 62:\penalty0 237--260, 01 2006.
\newblock \doi{10.2528/PIER06040901}.

\bibitem[200(2002)]{200289}
5 - radiation.
\newblock In Ron Schmitt, editor, \emph{Electromagnetics Explained}, EDN Series for Design Engineers, pages 89--109. Newnes, Burlington, 2002.
\newblock ISBN 978-0-7506-7403-4.
\newblock \doi{https://doi.org/10.1016/B978-075067403-4/50006-9}.
\newblock URL \url{https://www.sciencedirect.com/science/article/pii/B9780750674034500069}.

\bibitem[Papas(1954)]{Caltech}
Charles~H. Papas.
\newblock A note concerning a gyroelectric medium. caltech antenna laboratory technical report, 4. (unpublished).
\newblock \emph{California Institute of Technology , Pasadena, CA.}, \penalty0 (4), 1954.

\bibitem[Shastri et~al.(2021)Shastri, Abdelrahman, and Monticone]{photonics8040133}
Kunal Shastri, Mohamed~Ismail Abdelrahman, and Francesco Monticone.
\newblock Nonreciprocal and topological plasmonics.
\newblock \emph{Photonics}, 8\penalty0 (4), 2021.
\newblock ISSN 2304-6732.
\newblock \doi{10.3390/photonics8040133}.
\newblock URL \url{https://www.mdpi.com/2304-6732/8/4/133}.

\bibitem[Gibson(2008)]{bookmom}
Walton Gibson.
\newblock \emph{The Method of Moments in Electromagnetics}, volume~1.
\newblock 01 2008.
\newblock ISBN 9780429159046.
\newblock \doi{10.1201/b17119}.

\bibitem[Harrington(1982)]{harrington1982field}
R.F. Harrington.
\newblock \emph{Field Computation by Moment Methods}.
\newblock Macmillan Series in Electrical Science. R.E. Krieger Publishing Company, 1982.
\newblock ISBN 9780898744651.
\newblock URL \url{https://books.google.ps/books?id=-ANRAAAAMAAJ}.

\bibitem[Anastassiu(2006)]{1603812}
H.T. Anastassiu.
\newblock Fast, simple and accurate computation of the currents on an arbitrarily large circular loop antenna.
\newblock \emph{IEEE Transactions on Antennas and Propagation}, 54\penalty0 (3):\penalty0 860--866, 2006.
\newblock \doi{10.1109/TAP.2006.869929}.

\bibitem[Huba(2004)]{axcv}
Joseph Huba.
\newblock Nrl: Plasma formulary.
\newblock page~73, 12 2004.

\end{thebibliography}






\end{document}